\documentclass{ws-rmp}
\begin{document}

\title{Abelian gerbes, generalized geometries and foliations of small exotic $\mathbb{R}^{4}$}

\author{Torsten Asselmeyer-Maluga}

\address{German Aero space center, Rosa-Luxemburg-Str. 2, 10178 Berlin,
Germany \\ torsten.asselmeyer-maluga@dlr.de} 

\author{Jerzy Kr{\'o}l}

\address{University of Silesia, ul. Uniwesytecka 4, 40-007 Katowice,
Poland \\ iriking@wp.pl} 

\maketitle

\begin{history}
\received{(5 September 2012)}
\revised{(27 August 2014)}
\end{history}

\begin{abstract}
In the paper we prove the existence of the strict but relative relation
between small exotic $\mathbb{R}^{4}$ for a fixed radial family of
DeMichelis-Freedman type, and cobordism classes of codimension one
foliations of $S^{3}$ distinguished by the Godbillon-Vey invariant,
$GV\in H^{3}(S^{3},\mathbb{R})$ (represented by a 3-form). This invariant
can be integrated to get the Godbillon-Vey number. For a fixed radial
family, we will show that the isotopy classes (invariance w.r.t. small
diffeomorphisms or coordinate transformations) of all members in this
family are distinguished by the Godbillon-Vey number of the foliation
which is equal to the square of the radius of the radial family. The
special case of integer Godbillon-Vey invariants $GV\in H^{3}(S^{3},\mathbb{Z})$
is also discussed and is connected to flat $PSL(2,\mathbb{R})-$bundles.
Next we relate these distinguished small exotic smooth $\mathbb{R}^{4}$'s
to twisted generalized geometries of Hitchin on $TS^{3}\oplus T^{\star}S^{3}$
and abelian gerbes on $S^{3}$. In particular the change of the smoothness
on $\mathbb{R}^{4}$ corresponds to the twisting of the generalized
geometry by the abelian gerbe. We formulate the localization principle
for exotic 4-regions in spacetime and show that the existence of these
domains causes the quantization of electric charge, the effect usually
ascribed to the existence of magnetic monopoles. 
\end{abstract}

\ccode{PACS: 02.40.Sf, 02.40.Vh, 04.20.Gz}
\tableofcontents{}

\section{Introduction }

General relativity (GR) has changed our understanding of space-time.
In parallel, the appearance of quantum field theory (QFT) has modified
our view of particles, fields and the measurement process. The usual
approach for the unification of QFT and GR, to a quantum gravity,
starts with a proposal to quantize GR and its underlying structure,
space-time. There is a unique opinion in the community about the relation
between geometry and quantum theory: The geometry as used in GR is
classical and should emerge from a quantum gravity in the limit (Planck's
constant tends to zero). Most theories went a step further and try
to get a space-time from quantum theory. Then, the model of a smooth
manifold is not suitable to describe quantum gravity. But, there is
no sign for a discrete space-time structure or higher dimensions in
current experiments. Therefore, we conjecture that the model of spacetime
as a smooth 4-manifold can be used also in a quantum gravity regime.
But then one has the problem to represent QFT by geometric methods
(submanifolds for particles or fields etc.) as well to quantize GR.
Here, the exotic smoothness structure of 4-manifolds can help to find
a way. A lot of work was done in the last decades to fulfill this
goal. It starts with the work of Brans and Randall \cite{BraRan:93}
and of Brans alone \cite{Bra:94a,Bra:94b,Bra:99} where the special
situation in exotic 4-manifolds (in particular the exotic $\mathbb{R}^{4}$)
was explained. One main result of this time was the {\em Brans conjecture}:
exotic smoothness can serve as an additional source of gravity. It
was confirmed for compact manifolds by Asselmeyer \cite{Ass:96} and
for the exotic $\mathbb{R}^{4}$ by S{\l{}}adkowski \cite{Sla:99,Sladkowski2001}.
But this conjecture was extended in \cite{AssBra:2002} to conjecture
the generation for all forms of known energy, especially dark matter
and dark energy. For dark energy we were partly successful in \cite{AsselmeyerKrol2014}
where we calculated the expectation value of an embedded surface.
This value showed an inflationary behavior and we were also able to
calculate a cosmological constant having a realistic value (in agreement
with the Planck satellite results).

The inclusion of QFT was also another goal of our approach. Here we
will show that an exotic 4-manifold (and therefore the spacetime)
has a complicated foliation. Using noncommutative geometry, we were
able to study these foliations and got relations to QFT. For instance,
the von Neumann algebra of a codimension-one foliation of an exotic
$\mathbb{R}^{4}$ must contain a factor of type $I\! I\! I_{1}$ used
in local algebraic QFT to describe the vacuum\cite{AsselmeyerKrol2010,AsselmeyerKrol2011a,AsselmeyerKrol2011d}.
But why is an exotic 4-manifold so complicated? As an example let
us consider the exotic $S^{3}\times\mathbb{R}$. Clearly, there is
always a topologically embedded 3-sphere but there is no smoothly
embedded one. Let us assume the well-known hyperbolic metric of the
spacetime $S^{3}\times\mathbb{R}$ using the trivial foliation into
leafs $S^{3}\times\left\{ t\right\} $ for all $t\in\mathbb{R}$.
Now we demand that $S^{3}\times\mathbb{R}$ carries an exotic smoothness
structure at the same time. Then we will get only topologically embedded
3-spheres, the leafs $S^{3}\times\left\{ t\right\} $ (otherwise one
obtains the standard smoothness structure, see \cite{Chernov2012}
for instance). These topologically embedded 3-spheres are also known
as wild 3-spheres. In \cite{AsselmeyerKrol2011c}, we presented a
relation to quantum D-branes. Finally we proved in \cite{AsselmeyerKrol2013}
that the deformation quantization of a tame embedding (the usual embedding)
is a wild embedding. Furthermore we obtained a geometric interpretation
of quantum states: wild embedded submanifolds are quantum states.
Importantly, this construction depends essentially on the continuum,
wild embedded submanifolds admit always infinite triangulations. This
approach opens a way to quantize a theory using geometric methods.

The inclusion of matter is also one of the main problems in this theory.
But after the confirmation of the Brans conjecture, we supposed that
there must be one way to introduce it. For a special class of compact
4-manifolds we showed in \cite{AsselmeyerRose2012} that exotic smoothness
can generate fermions and gauge fields using the so-called knot surgery
of Fintushel and Stern \cite{FinSte:98}. Here, the knot is directly
related to the appearance of an exotic smoothness structure. Then
one obtains a stable but fixed structure of fermions and gauge fields
contradicting the results of QFT with a variable number of fermions
and gauge fields%
\footnote{In the following we don't make a difference between a fermionic quantum
field and a fermion.%
} (where the (virtual) fermions and gauge fields will be generated
or destroyed). In the paper \cite{AsselmeyerBrans2014} we presented
an approach using the exotic $\mathbb{R}^{4}$ solving the difficulties
of \cite{AsselmeyerRose2012}. The special role of the exotic $\mathbb{R}^{4}$
is given by the fact (for all known exotic $\mathbb{R}^{4}$) that
the neighborhood of every compact subset in the exotic $\mathbb{R}^{4}$
is surrounded by a compact 3-manifold (not homeomorphic to the 3-sphere).
Therefore we obtain always a non-trivial 3-manifold from an exotic
$\mathbb{R}^{4}$ whereas for the standard $\mathbb{R}^{4}$ one will
always find a neighborhood which is surrounded by a 3-sphere. But
this non-trivial 3-manifold (inside the exotic $\mathbb{R}^{4}$)
is not uniquely determined, it depends on the representation of the
exotic $\mathbb{R}^{4}$ and on the choice of the neighborhood. From
the physical point of view, this behavior is known from QFT where
a fermion is surrounded by a 'cloud' of virtual particles. In \cite{AsselmeyerRose2012}
we obtained a complete picture of known matter: fermions as hyperbolic
knot complements and gauge fields as torus bundles. This picture should
be also extended to the exotic $\mathbb{R}^{4}$ in our forthcoming
work. Furthermore the relation to quantum gravity has to be understand
more completely. First signs of a relation can be found in \cite{Pfeiffer2004,Duston2010,Ass2010}
or by using string theory \cite{AssKrol2010ICM}.

The main motivation for this paper came from the following problem.
Every 3-manifold has a unique smoothness structure in contrast to
a 4-manifold. Now let $M^{4}$ be a 4-manifold with an exotic smoothness
structure and boundary $\Sigma^{3}=\partial M^{4}$. The boundary
admits a unique smoothness. But how can one detect the exotic smoothness
structure in the interior? Which manifold structure on the boundary
changes if we change the smoothness structure in the interior? One
can call it: the \emph{holography problem for exotic 4-manifolds}.
In this paper we will present a solution to this problem for small
exotic $\mathbb{R}^{4}$: the codimension-one foliation of the boundary
at infinity will change! This fact has a tremendous impact on the
geometry. Furthermore, every small exotic $\mathbb{R}^{4}$, denoted
by $R^{4}$, embeds in the standard $\mathbb{R}^{4}$, in the following
denoted by $\mathbf{R}^{4}$. Then this foliation goes over to a 3-sphere
in $\mathbf{R}^{4}$ and can be detected by abelian gerbes and deformed
Hitchin structures. From the physical point of view, we will obtain
the quantization of the electrical charge by using $R^{4}$ inside
of $\mathbf{R}^{4}$ having the same effect as a magnetic monopole. 

Now we will remark about isotopy classes and physics. GR is invariant
w.r.t. coordinate transformations, also called small diffeomorphisms.
But there are also large diffeomorphisms. A simple example is given
by a Dehn twist of a torus: Cut the torus at one place (getting a
cylinder), twist one side by $2\pi$ and glue both sides together.
One obtains a twisted torus which is diffeomorphic to the original
one but only by a large diffeomorphism. The twisted torus is a non-trivial
isotopy class of a torus. But from the physical perspective, both
different isotopy classes are physically different (see also \cite{Giulini94}).
This simple example shows the importance of isotopy classes and goes
over to GR (on spacetime) as well. In particular, the topology of
the configuration space in canonical quantum gravity is strongly influenced
by the isotopy group of 3-manifold \cite{Witt1986}.

\subsection{Main ideas and motivation: exotic $\mathbb{R}^{4}$ and foliations}

Here we will give a short account of the main construction.

\subsubsection{The radial family of exotic $\mathbb{R}^{4}$}

\begin{figure}
\centerline{\psfig{file=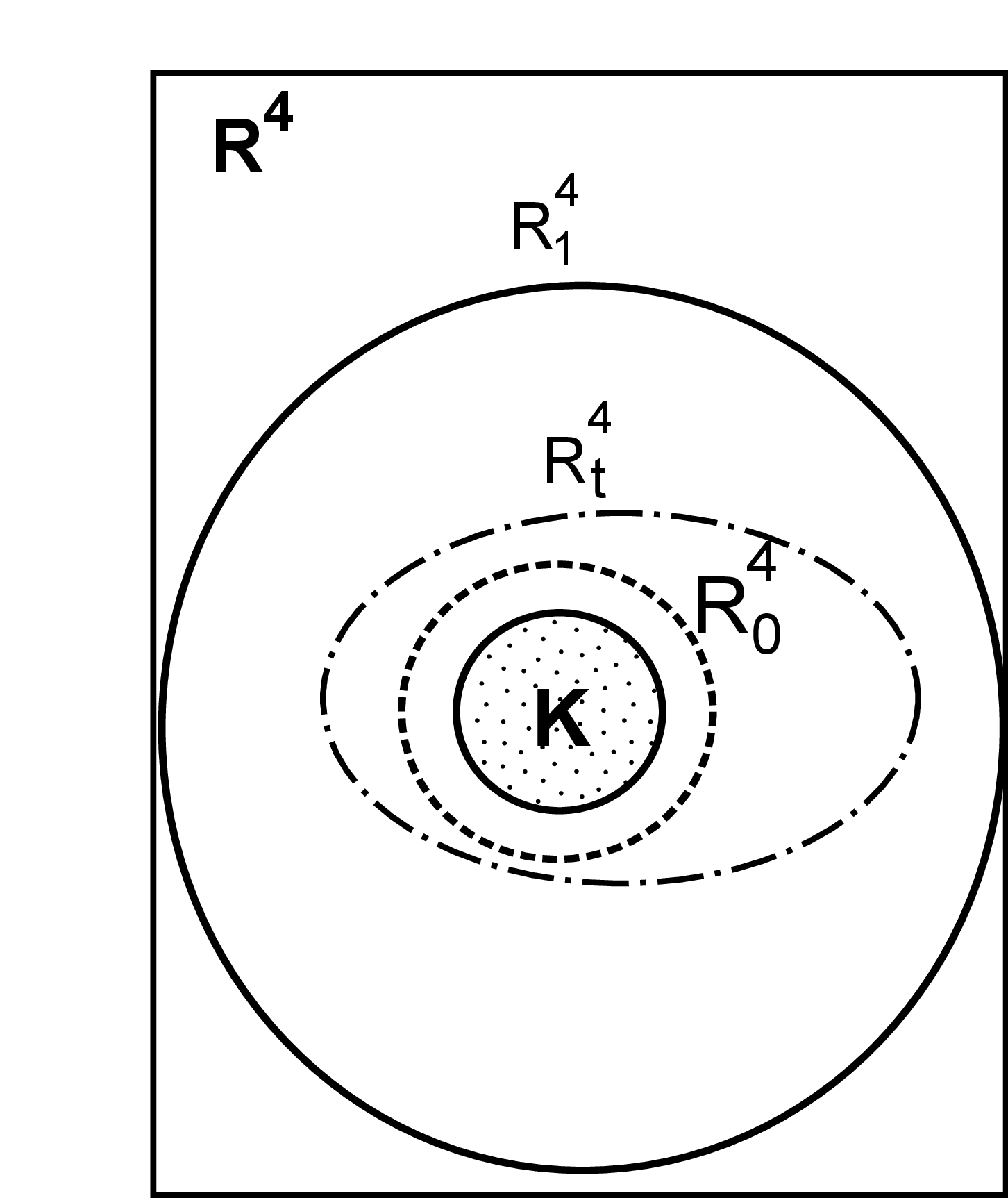,width=6cm}}
\caption{Radial family of small exotic $\mathbb{R}^{4}$ (with compact subset
$K$) parametrized by $[0,1]$ and embedded in the standard $\mathbb{R}^{4}$\label{fig:Radial-family-exoticR4}}
\end{figure}

Starting point is a small exotic $\mathbb{R}^{4}$, called $R^{4}$,
i.e. an open, noncompact 4-manifold homeomorphic to $\mathbb{R}^{4}$
having the standard smoothness structure (so that $\mathbb{R}^{4}$
is smoothly splittable $\mathbb{R}^{3}\times\mathbb{R}$) but not
diffeomorphic to it and can be embedded in the 4-sphere $S^{4}$ (in
contrast to a large exotic $\mathbb{R}^{4}$ which fails to have this
embedding). One of the characterizing properties of an exotic $\mathbb{R}^{4}$
(all known examples) is the existence of a compact subset $K\subset R^{4}$
which cannot be surrounded by any smoothly embedded 3-sphere (and
homology 3-sphere bounding a contractable, smooth 4-manifold). This
property will be used to construct the radial family of uncountably
many, different exotic $\mathbb{R}^{4}$. Let $\mathbf{R}^{4}$ be
the standard $\mathbb{R}^{4}$ (i.e. $\mathbf{R}^{4}=\mathbb{R}^{3}\times\mathbb{R}$
smoothly) and let $R^{4}$ be an exotic $\mathbb{R}^{4}$ with compact
subset $K\subset R^{4}$ which cannot be surrounded by a smoothly
embedded 3-sphere. Now fix a homeomorphism 
\[
h:\mathbf{R}^{4}\to R^{4}
\]
and define the image 
\[
R_{r}^{4}=h(B_{r}^{4})
\]
of an open ball $B_{r}^{4}=\left\{ x\in\mathbf{R}^{4}|\,||x||^{2}<r\right\} \subset\mathbf{R}^{4}$
of radius $r\in[0,\infty]$ centered at $0$ in $\mathbf{R}^{4}$.
We call $R_{r}^{4}$ the \emph{radial family} with $R_{\infty}^{4}=R^{4}$
(or $R_{\infty}^{4}=\mathbf{R}^{4}$ the standard $\mathbb{R}^{4}$).
See Fig. \ref{fig:Radial-family-exoticR4} for the visualization of
a radial family of small exotic $\mathbb{R}^{4}$. 

Of course, there is a minimal radius $r_{0}$ so that $K\subset R_{r_{0}}^{4}$and
then $R_{r_{0}}^{4}\subset R^{4}$ is an exotic $\mathbb{R}^{4}$
(Later we identify $R_{r_{0}}^{4}$with $R_{0}^{4}$.). For different
radii, say $s$ and $t$, one can show that $R_{s}^{4}$ and $R_{t}^{4}$
are mostly (up to countable many pairs of values) not diffeomorphic
to each other (see \cite{Gom:85,Tau:87} for large exotic $\mathbb{R}^{4}$
and \cite{DeMichFreedman1992} for small exotic $\mathbb{R}^{4}$).
Furthermore, for $s<t$ one has $R_{s}^{4}\subset R_{t}^{4}$ so that
every member in this family is an exotic $\mathbb{R}^{4}$(see also
\cite{GomSti:1999} sec 9.4). We also remark that the smoothly embedded
3-manifold which surrounds the compact subset $K$ is an expression
for the complexity of the exotic $\mathbb{R}^{4}$ (see \cite{Ganzel2006}).
Secondly, we remark that $K$ is of course surrounded by a topological
3-sphere but this topological 3-sphere is wildly embedded. Wild embedded
manifolds (like Alexanders horned sphere or the Fox-Artin wild knot)
are characterized by the property that a wild embedded manifold must
be triangulated by infinite many polyhedrons. If $R^{4}$ is a small
exotic $\mathbb{R}^{4}$ then there is a smooth embedding $R^{4}\to\mathbf{R}^{4}$
in the standard $\mathbb{R}^{4}$.

\subsubsection{Motivation: Foliations of exotic $\mathbb{R}^{4}$}

All exotic $\mathbb{R}^{4}$ admit complicated foliations which will
be explained now. Usually there are arbitrary splittings of the $\mathbb{R}^{4}$
into $\mathbb{R}^{3}\times\mathbb{R}$ or $\mathbb{R}\times\mathbb{R}\times\mathbb{R}\times\mathbb{R}$.
But every manifold of dimension smaller than 4 has an unique smoothness
structure. Therefore $\mathbb{R}^{3}\times\left\{ t\right\} $ has
an unique smoothness structure for every $t\in\mathbb{R}$ and also
the whole $\mathbb{R}^{3}\times\mathbb{R}$. A standard smoothness
structure on $\mathbb{R}^{4}$ is uniquely characterized by the fact
that the smoothness structure respects the decomposition $\mathbb{R}^{3}\times\mathbb{R}^{1}$.
But more is true: also every splitting $\Sigma\times\mathbb{R}$ with
the contractable 3-manifold $\Sigma$ is diffeomorphic to $\mathbf{R}^{4}$
(standard $\mathbb{R}^{4}$), see \cite{Chernov2012}. Expressed in
physical terms: there is no globally hyperbolic metric on any exotic
$\mathbb{R}^{4}$. But by using foliation theory, every $\mathbb{R}^{4}$
admits a codimension-one foliation (or there is a non-vanishing vector
field which can be used to define a Lorentz metric, see \cite{Ste:99}).
Therefore there are close connections between exotic $\mathbb{R}^{4}$
and non-trivial codimension-one foliations.

In the concrete example of a small exotic $\mathbb{R}^{4}$, we are
able to construct this foliation for a fixed radial family. At first
we will discuss the type of the foliation. Small exotic $\mathbb{R}^{4}$'s
were constructed by using the failure of the smooth h-cobordism theorem.
Let $M,M_{0}$ be a compact, closed and simply-connected 4-manifolds
together with a h-cobordism $W$ (i.e. a 5-manifold $W$ with $\partial W=M\sqcup M_{0}$
and the embeddings $M,M_{0}\hookrightarrow W$ induce homotopy-equivalences).
Then $M$ and $M_{0}$ are homeomorphic \cite{Fre:82} and there are
contractable submanifolds $A_{0}\subset M_{0},\,\, A\subset M$ and
a h-subcobordism $X\subset W$ with $\partial X=A_{0}\sqcup A$. As
shown in \cite{CuFrHsSt:97}, the remaining h-cobordism $W\setminus X$
trivializes $W\setminus X=(M_{0}\setminus A_{0})\times[0,1]$ inducing
a diffeomorphism between $M_{0}\setminus A_{0}$ and $M\setminus A$.
The submanifolds $A,A_{0}$ are called Akbulut corks. If $M$ and
$M_{0}$ are not diffeomorphic then the open neighborhoods $U,U_{0}$
of $A$ and $A_{0}$, respectively, are diffeomorphic small exotic
$\mathbb{R}^{4}$'s. One possible way to measure the difference between
$M$ and $M_{0}$ is given by the Seiberg-Witten invariant. At first
we note that the difference between $M$ and $M_{0}$ is given by
the embedding of the Akbulut cork resulting in the submanifolds $A$
and $A_{0}$, respectively. Let us assume that $M_{0}$ carries a
standard smoothness structure. Then the Seiberg-Witten invariant of
$M_{0}$ is zero whereas the Seiberg-Witten invariant of $M$ is non-zero.
But then the solution of the Seiberg-Witten equations have to be non-zero
which implies the existence of a submanifold with negative scalar
curvature (see \cite{Lebrun96,Lebrun98}). By the argumentation above,
this submanifold must be located at the open neighborhood $U$ of
$A$, or the small exotic $\mathbb{R}^{4}$ has to carry a geometry
with negative scalar curvature (for instance a hyperbolic geometry).
Foliations implying negative scalar curvature are known to have at
least the group $PSL(2,\mathbb{R})$ as isometry group, the simplest
isometry group of a hyperbolic space. An example of this foliation
was given by Thurston \cite{Thu:72}. Starting point of the construction
is a convex polygon $K$ with $q$ vertices in the hyperbolic space
$\mathbb{H}^{2}$ which will be doubled along one axis to get a sum
of two polygons $K\cup K'$. Small neighborhoods $U(p)$ of the vertices
$p$ will be removed from the polygons. Using the isometry group $PSL(2,\mathbb{R})$
of $\mathbb{H}^{2}$, one can identify the corresponding sides of
the polygons $K$ and $K'$ to get a compact 2-manifold $V^{2}$ diffeomorphic
to a 2-sphere with $q$ holes. This space $V^{2}$ can be foliated
(with respect to the isometry group $PSL(2,\mathbb{R})$) which can
be extended to the unit tangent bundle $U\mathbb{H}^{2}$ (the tangent
bundle of unit vectors) diffeomorphic to $V^{2}\times S^{1}$. The
$q$ holes can be filled by gluing in copies of $D^{2}\times S^{1}$
which are foliated with a Reeb foliation. By using the Dehn-Lickorish
theorem (see \cite{PrasSoss:97} Corollary 12.4), one obtains every
possible compact 3-manifold for different choices of the gluing map
for the copies $D^{2}\times S^{1}$, among them the 3-sphere. The
Godbillon-Vey invariant of this foliation is a 3-form which can be
integrated by using the fundamental class of the 3-manifold to get
a Godbillon-Vey number. Therefore we motivated a relation between
codimension-one foliations, small exotic $\mathbb{R}^{4}$ and hyperbolic
geometry.

\subsubsection{Main construction}

This subsection contains the main ideas of the construction. At first
we will express the idea in (more or less) non-technical terms: 
\begin{enumerate}
\item Every member of the radial family $R_{t}^{4}$ is determined by a
Casson handle represented by a rooted tree (see \cite{GomSti:1999}
Theorem 9.4.12). The members are ordered w.r.t. the parameter $t$,
i.e. $R_{0}^{4}\subset R_{s}^{4}\subset R_{t}^{4}\subset R_{1}^{4}$
for $s<t$ (see Fig. \ref{fig:Radial-family-exoticR4}). The completion
of one member $\overline{R_{t}^{4}}$ has a boundary $Y_{r}^{3}=\partial\overline{R_{t}^{4}}$
(where the properties of $R_{t}^{4}$ go over). This 3-manifold $Y_{r}^{3}$
admits a codimension-one foliation. The idea of this construction
uses the description of $Y_{r}^{3}$ by knots/links (via surgery \cite{GomSti:1999}).
Loosely speaking, the knot/link induces this foliation. 
\item The whole family admits a foliation induced by the foliation of every
member. An important point in this construction is the appearance
of rigidity, i.e. $Y_{r}^{3}$ contains hyperbolic 3-manifolds of
finite volume. But a hyperbolic 3-manifold of finite volume does not
scale (Mostow rigidity), i.e. a diffeomorphism is an isometry. This
property implies the fixing of the size of a disk and (using the construction
above) the fixing of the Godbillon-Vey number. 
\item Furthermore the appearance of hyperbolic 3-manifolds guarantees that
the foliation is $PSL(2,\mathbb{R})$ invariant and has a non-zero
Godbillon-Vey number \cite{ReinhartWood1973}. Finally this foliation
has a non-zero Godbillon-Vey number which is determined by the radius
of the member $R_{t}^{4}$ in the radial family to be $GV=r^{2}$. 
\end{enumerate}
Now we will give a short overview of technical details. Let $\mathbf{R}^{4}$
be the standard and $R^{4}$ be an exotic $\mathbb{R}^{4}$ of small
type. We fix a homeomorphism $h:\mathbf{R}^{4}\to R^{4}$. As shown
in \cite{FreQui:90}, every homeomorphism between smooth 4-manifolds
is isotopic to one which is a local diffeomorphism near a preassigned
1-complex. In our case it means that one can introduce a topological
radius function (polar coordinate) $\rho:R^{4}\to[0,+\infty)$ defining
this 1-complex for the smoothing of the homeomorphism $h$. According
to \cite{DeMichFreedman1992} Theorem 3.2, if we set $t=1-1/r$ and
$R_{t}^{4}=\rho^{-1}([0,r))$ so that $R_{1}^{4}$ is a small exotic
$\mathbb{R}^{4}$ containing the compactum $K\subset R_{1}^{4}$ then
$K\subset R_{0}^{4}$ ($t=0$) and $R_{t}^{4}$ is also a small exotic
$\mathbb{R}^{4}$ for every $t$ belonging to the Cantor set $CS\subset[0,1]$
(so that the radius $r$ is in the range $[1,+\infty)$). The definition
of the radial family implies that $K\subset R_{t}^{4}$ for every
$t$ so that every $R_{t}^{4}$ is a small exotic $\mathbb{R}^{4}$
but all different (non-diffeomorphic) small exotic $\mathbb{R}^{4}$
are parametrized by the Cantor set $CS$ modulo a countably subset
of isotopy classes (see below). Every member $R_{t}^{4}$ of the radial
family has the following property: there is an open neighborhood $U(K)$
of the compactum $K\subset R_{t}^{4}$ having the 3-manifold $\partial\overline{U(K)}$
as boundary of the completion $\overline{U(K)}$. The 3-manifold $\partial\overline{U(K)}$
separates $K$ from infinity (see \cite{Ganzel2006}). By the topological
radius function $\rho$ we have also a compact space $B_{t}^{4}=\rho^{-1}([0,r])$
with a boundary $Y=\partial B_{t}^{4}$. Of course $Y$ separates
$K$ from infinity in $R_{t_{1}}^{4}$ (i.e. for a larger $r_{1}>r$
with $t_{1}=1-1/r_{1}$). Obviously, the size of $Y$ scales with
the radius $r$ like $vol(Y)\sim r^{3}$ and we will denote it by
$Y_{r}$. Of course for a concrete example, one can determine $Y_{r}$
but we will be as general as possible. Therefore we have to ask about
a common submanifold of all $Y_{r}$. Here we will use the Dehn-Lickorish
theorem, i.e. every 3-manifold can be decomposed as 
\begin{equation}
Y_{r}=\left(V^{2}\times S^{1}\right)\cup_{\phi_{1}}\left(D^{2}\times S^{1}\right)\cup_{\phi_{2}}\cdots\cup_{\phi_{q}}\left(D^{2}\times S^{1}\right)\label{eq:decomposition-3MF-Y}
\end{equation}
by choosing a large enough number $q\in\mathbb{N}$, where $V^{2}=S^{2}\setminus\left(\bigsqcup^{q}D^{2}\right)$
(see above) and $\phi_{n}:S^{1}\times S^{1}\to S^{1}\times S^{1}$
as Dehn twists representing the gluing map. The specific topology
of every 3-manifold is encoded into the choice of the gluing maps
$\phi_{n}$. Furthermore, the representation of the 3-manifold $Y_{r}$
is not unique. The transformation rules from one representation to
a new one are called Kirby calculus (see \cite{GomSti:1999} or the
original papers \cite{Kir:78,FenRou:79}). Now we are ready to present
our main arguments: 
\begin{enumerate}
\item The 3-manifold $Y_{r}$ is given by the radius of the member $R_{t}^{4}$
(with $t=1-1/r$) which determines also the size of the components
in the decomposition (\ref{eq:decomposition-3MF-Y}). We can choose
all $D^{2}\times S^{1}$ to be small whereas $vol(V^{2}\times S^{1})\sim r^{3}$
scales like $Y_{r}$. Then we obtain also $vol(V^{2})\sim r^{2}$.
Fixing a member of the radial family is equal to fixing the size of
$Y_{r}$ and $V^{2}$. But $V^{2}$ is topologically equivalent to
the complex $K\cup K'\setminus U(p_{1},\ldots,p_{q})$ in $\mathbb{H}^{2}$
(by the Riemann uniformization theorem). Fixing the size of $V^{2}$
(up to a scaling of the whole radial family) determines the transformation
group of $V^{2}$ to be $PSL(2,\mathbb{R})$. A unit sphere bundle
over $\mathbb{H}^{2}$ defines a foliation of $V^{2}\times S^{1}$
(by geodesics) with Godbillon-Vey number proportional to $vol(V^{2})$,
i.e. to $r^{2}$. 
\item The next argument is more sophisticated and uses the details of the
proof in \cite{DeMichFreedman1992} to show the existence of a foliation.
Every member $R_{t}^{4}$ of the radial family depends on the Casson
handle, i.e. an open 2-handle homeomorphic to $D^{2}\times\mathbb{R}^{2}$
but represented by a rooted tree. At first we will construct a foliation
at $Y_{r}$ which depends only on the Casson handle (Theorem \ref{thm:foliation-single-small-exotic-R4})
using Gabai's work \cite{Gabai1983} and the splitting of the knot/link
complement \cite{Budney2006}. Using Mostow rigidity, the foliation
is rigid (Lemma \ref{lem:size-control-Y} and Lemma \ref{lem:foliation-Y-for-member}
also appendix B) i.e. a small diffeomorphism (or coordinate transformation)
does not change it. The rigidity determines also the size of a polygon
(or disk) in the complement. The 3-manifold $Y_{r}$ is compact and
therefore the tree representing the Casson handle has to embed in
a compact manifold. All trees are planar and we can choose a disk
of fixed radius $r$ and area $r^{2}$. Then the Godbillon-Vey number
is proportional to $r^{2}$. 
\item For the construction of the whole radial family one needs a parametrization
of all Casson handles (from a given Casson handle), the so-called
design. But this description is not complete, i.e. there are Casson
handles which are not embeddable in the given Casson handle or containing
the given Casson handle as embedding, called gaps. The design admits
a foliation transversal to $Y_{r}$ induced by a foliation on $Y_{r}$.
Then using the lemma \ref{lem:foliation-gap-1}, we will obtain a
foliation on the gaps too. 
\item Then we will put all arguments together in Theorem \ref{thm:codim-1-foli-radial-fam}.
But the whole construction depends essentially on the Casson handle.
Currently there is no complete classification of smooth Casson handles
but we know that two different Casson handle (having different rooted
trees as representatives) are non-isotopic to each other (see subsection
\ref{sub:Isotopy-of-surfaces}). Therefore our argumentation is mainly
via the isotopy classes and its relation to the foliation. Finally
(using argument 2 above) we obtain the Godbillon-Vey number to be
$r^{2}$. 
\end{enumerate}
In principle, the argumentation at the first item above is enough
to obtain a relation to the Godbillon-Vey invariant but the choice
of the size of $V^{2}$ seems arbitrary. Therefore we needed the advanced
results of the other item to get the full theorem. Via the embedding
of the member $R_{t}^{4}$ into $\mathbf{R}^{4}$ (because of the
smallness), one gets also a relation between the foliation of $Y_{r}$
and $S^{3}$. Let $U(R_{t}^{4})$ be a neighborhood of the image of
the embedding $I_{t}:R_{t}^{4}\hookrightarrow\mathbf{R}^{4}$ with
boundary $S^{3}=\partial U(R_{t}^{4})$ a 3-sphere. This 3-sphere
$S^{3}\subset\mathbf{R}^{4}$ admits also a decomposition 
\[
S^{3}=\left(V_{S^{3}}^{2}\times S^{1}\right)\cup_{\phi_{1}}\left(D^{2}\times S^{1}\right)\cup_{\phi_{2}}\cdots\cup_{\phi_{q}}\left(D^{2}\times S^{1}\right)
\]
(where $\phi_{i}$ is the identity except for one map, say $\phi_{1}$)
so that we are able to define a foliated cobordism $B(V_{S^{3}}^{2}\times S^{1},V^{2}\times S^{1})$
between $V_{S^{3}}^{2}\times S^{1}$ and $V^{2}\times S^{1}$. Then
the foliation of $V^{2}\times S^{1}$ induces a foliation of $V_{S^{3}}^{2}\times S^{1}$.
The Godbillon-Vey invariant is an invariant of the foliated cobordism,
i.e. the (induced) foliation of the 3-sphere has the same Godbillon-Vey
invariant like the foliation of $Y_{r}$.

\subsection{Organization of the Paper}

The paper is organized as follows. In the next section we present
a short introduction into codimension-one foliations on 3-manifolds
and isotopy of surfaces related to rooted trees. We will need this
material in the following sections. Then we will derive the relation
between exotic $\mathbb{R}^{4}$ and codimension-one foliations of
$S^{3}$ and obtain the main result (Theorem \ref{thm:codim-1-foli-radial-fam}
and \ref{thm:foliation-S3-from-Y}) of the paper: \\
 \emph{A fixed radial family $R_{t}^{4}$ of small exotic $\mathbb{R}^{4}$
with radius $r$ and $t=1-\frac{1}{r}\subset CS\subset[0,1]$ (induced
from a non-product h-cobordism $W$ between $M$ and $M_{0}$ with
Akbulut corks $A\subset M$ and $A\subset M_{0}$, respectively) contains
a compact 4-dimensional submanifold $K$ which cannot be surrounded
by a smoothly embedded 3-sphere. Then $K\subset R_{0}^{4}\subset R_{1}^{4}$
and every two members $R_{s}^{4},R_{t}^{4}$ with $s\not=t\in[0,1]$
are non-isotopic and some ($s,t\in CS$) are non-diffeomorphic to
each other.} \emph{Every member $R_{s}^{4}\subset R_{t}^{4}$ with
$s<t$ embeds and there is a neighborhood $U(R_{s}^{4})\subset R_{t}^{4}$
with boundary $Y_{s}=\partial U(R_{s}^{4})$ a 3-manifold.} \emph{Every
member $R_{t}$ of the radial family determines a codimension-one
foliation of $Y_{r}$ with Godbillon-Vey number $r^{2}=\frac{1}{(1-t)^{2}}$}.\emph{
Let $R_{t}$ and $R_{s}$ with $s\not=t$ be two members, then the
two members are non-isotopic and the two corresponding codimension-one
foliations of $Y_{r}$ and $Y_{u}$, respectively, are non-cobordant
to each other. Furthermore every $R_{t}^{4}$ embeds into $\mathbf{R}^{4}$,
the standard $\mathbb{R}^{4}$, so that there is a neighborhood $U(I_{t}(R_{t}^{4}))$
of the embedding $I_{t}:R_{t}^{4}\hookrightarrow\mathbf{R}^{4}$ with
boundary a 3-sphere $S^{3}$. Then there is a foliated cobordism between
two topologically equivalent submanifolds of $S^{3}$ (in $\mathbf{R}^{4}$)
and $Y_{r}$ (of $R_{t}^{4}$) so that the induced foliations have
equal Godbillon-Vey numbers. }\\
 The standard $\mathbf{R}^{4}$ is not far from a small exotic $R^{4}$.
In Sec. \ref{sec:deformation-standard-R4} we will discuss the deformation
of $\mathbf{R}^{4}$ to get $R^{4}$ by using a wild embedded submanifold,
the Whitehead continuum. The effect of this deformation on the function
algebra, the geometry (holonomy) and vector fields is also discussed.
This section serves as a motivation for the introduction of abelian
gerbes and Hitchin structures in the following.

The Bockstein homomorphism (induced by the injective map $\mathbb{Z}\to\mathbb{R}$)
defines integer GV numbers. Then we will give a direct characterization
of the codimension-one foliations of $S^{3}$ with integer GV number
(Theorem \ref{thm:integer-GV-flat-bundles}): These foliations correspond
to the flat $PSL(2,\mathbb{R})$ bundles on $M=S^{2}\setminus\left\{ k-{\rm punctures}\right\} \times S^{1}$
and the GV invariant of this foliation (when evaluated on the fundamental
class of $M$) is equal $\pm(2-k)$. Next in Sec. \ref{sub:U(1)--gerbes-on}
we will present a short introduction into abelian gerbes and the relation
of the exotic $\mathbb{R}^{4}$ (derived from the codimension-one
foliations of $S^{3}$ with integer GV classes) to abelian gerbes
on the 3-sphere $S^{3}$. In Sec. \ref{sub:Generalized-complex-str}
we are able to assign the changes of smoothness on $\mathbb{R}^{4}$
to twistings of the generalized geometries on $TS^{3}\oplus T^{\star}S^{3}$
by abelian gerbes. Based on these results we will show in the last
section that the appearance of small exotic $\mathbb{R}^{4}$ in 4-spacetime
can be used to explain the quantization of the electrical charge but
without magnetic monopoles.

\section{Preliminaries: Foliations on 3-manifolds and Isotopies of Surfaces}

In short, a foliation of a smooth manifold $M$ is an integrable subbundle
$N\subset TM$ of the tangent bundle $TM$. The existence of codimension-one-foliations
depends strongly on the compactness or non-compactness of the manifold.
Every compact manifold admits a codimension-one-foliation if and only
if the Euler characteristics vanish. In the following we will first
concentrate on the 3-sphere $S^{3}$ with vanishing Euler characteristics
admitting codimension-one-foliations.

\subsection{Definition of foliation and foliated cobordism\label{sub:Definition-of-Foliation-cobordism}}

A codimension $k$ foliation of an $n$-manifold $M^{n}$ (see the
nice overview article \cite{Law:74}) is a geometric structure which
is formally defined by an atlas $\left\{ \phi_{i}:U_{i}\to M^{n}\right\} $,
with $U_{i}\subset\mathbb{R}^{n-k}\times\mathbb{R}^{k}$, such that
the transition functions have the form 
\[
\phi_{ij}(x,y)=(f(x,y),g(y)),\,\left[x\in\mathbb{R}^{n-k},y\in\mathbb{R}^{k}\right]\quad.
\]
For a precise definition see the book \cite{Tamura1992}. Intuitively,
a foliation is a pattern of $(n-k)$-dimensional stripes - i.e., submanifolds
- on $M^{n}$, called the leaves of the foliation, which are locally
well-behaved. The tangent space to the leaves of a foliation $\mathcal{F}$
forms a vector bundle over $M^{n}$, denoted $T\mathcal{F}$. The
complementary bundle $\nu\mathcal{F}=TM^{n}/T\mathcal{F}$ is the
normal bundle of $\mathcal{F}$. Such foliations are called regular
in contrast to singular foliations or Haefliger structures. For the
important case of a codimension-one foliation we need an overall non-vanishing
vector field or its dual, an one-form $\omega$. This one-form defines
a foliation iff it is integrable, i.e. 
\begin{equation}
d\omega\wedge\omega=0\label{eq:integrability-foliation}
\end{equation}
The cross-product $M\times N$ defines a trivial foliation. One of
the first examples of a nontrivial foliation is known as Reeb foliation
(see \cite{Tamura1992}). Now we will discuss an important equivalence
relation between foliations, cobordant foliations. Let $M_{0}$ and
$M_{1}$ be two closed, oriented $m$-manifolds with codimension-$q$
foliations. Then these foliated manifolds are said to be \emph{foliated
cobordant} if there is a compact, oriented $(m+1)$-manifold with
boundary $\partial W=M_{0}\sqcup\overline{M}_{1}$ and with a codimension-$q$
foliation $\mathcal{F}$ transverse to the boundary. Every leaf $L_{\alpha}$
of the foliation $\mathcal{F}$ induces leafs $L_{\alpha}\cap\partial W$
of foliations $\mathcal{F}_{M_{0}},\mathcal{F}_{M_{1}}$on the two
components of the boundary $\partial W$ (see \cite{Tamura1992} \S29).

\begin{figure}
\centerline{\psfig{file=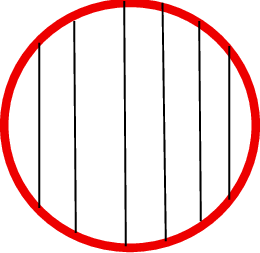,width=3cm}}
\caption{foliation of the disk $D^{2}$ \label{fig:foliation-of-the-disk}}
\end{figure}

As an example we consider a disk $D^{2}$ described by a complex number
$z=x+iy$ with $|z|\leq1$ and center $0$ together with a foliation
by leafs $L_{x}=\left\{ z=x+iy\in\mathbb{C}|\:-\frac{1}{2}\leq x\leq\frac{1}{2},\:|y|=\sqrt{1-x^{2}}/2\right\} $
(see Fig. \ref{fig:foliation-of-the-disk}). 

The intersections $L_{x}\cap\partial D^{2}$ of the leafs with the
boundary induces a foliation of $\partial D^{2}=S^{1}$ (the leafs
are only points). In this example we can also reverse this process,
i.e. given a foliation of $S^{1}$ by points we will obtain a foliation
of the disk $D^{2}$ by connecting the two points $(x,\sqrt{1-x^{2}}/2)$
and $(x,-\sqrt{1-x^{2}}/2)$ by a geodesic line.

As an example of a foliated cobordism we consider the foliation of
the cylinder $S^{1}\times[0,1]$ by the leafs $L_{\psi}=\left\{ (\psi,t)\in S^{1}\times[0,1]|\:0\leq t\leq1\right\} $
for $0\leq\psi\leq2\pi$. This foliation is obviously transverse to
the boundary and induces a foliation on every $S^{1}$ of the cobordism.
Now we close the cobordism by adding two disks on each side, i.e.
$D^{2}\cup(S^{1}\times[0,1])\cup D^{2}=S^{2}$ to get a 2-sphere with
the corresponding foliation (see Fig. \ref{fig:foliation-of-S2-1}).
This foliation of the 2-sphere gives the obvious foliation of the
3-disk $D^{3}$ (with $\partial D^{3}=S^{2}$) like in the case of
the disk above. But this foliation of the 3-disk can be understood
as a foliated cobordism between two disks forming the $S^{2}$ as
the boundary of the 3-disk $D^{3}$. We will later use this example
in the proof of our main theorem.

\subsection{Non-cobordant foliations of $S^{3}$ detected by the Godbillon-Vey
class \label{sub:Non-cobordant-foliations-S3}}

In \cite{Thu:72} Thurston constructed a foliation of the 3-sphere
$S^{3}$ which depends on a polygon $P$ in the hyperbolic plane $\mathbb{H}^{2}$
so that two foliations are non-cobordant if the corresponding polygons
have different areas. For later usage, we will present the main ideas
of this construction only (see also the book \cite{Tamura1992} chapter
VIII for the details). Starting point is the hyperbolic plane $\mathbb{H}^{2}$
with a convex polygon $K\subset\mathbb{H}^{2}$ having $k$ sides
$s_{1},\ldots,s_{k}$. Assuming the upper half plane model of $\mathbb{H}^{2}$
then the sides are circular arcs. The construction of the foliation
depends mainly on the isometry group $PSL(2,\mathbb{R})$ of $\mathbb{H}^{2}$
realized as rational transformations (and this group can be lifted
to $SL(2,\mathbb{R})$). The followings steps are needed in the construction: 
\begin{enumerate}
\item The polygon $K$ is doubled along one side, say $s_{1}$, to get a
polygon $K'$. The sides are identified by (isometric) transformations
$s_{i}\to s_{i}'$ (as elements of $SL(2,\mathbb{R})$). 
\item Take $\epsilon$-neighborhoods $U_{\epsilon}(p_{i}),U_{\epsilon}(p_{i}')$
with $\epsilon>0$ sufficient small and set 
\begin{align*}
V^{2} & =\left(K\cup K'\right)\setminus\bigcup_{i=1}^{k}\left(U_{\epsilon}(p_{i})\cup U_{\epsilon}(p_{i}')\right)\\
 & =S^{2}\setminus\bigcup_{i=1}^{k}D_{i}^{2}
\end{align*}
having the topology of $V^{2}=S^{2}\setminus\left\{ k\,\mbox{punctures}\right\} $
and we set $P=K\cup K'$. 
\item Now consider the unit tangent bundle $U\mathbb{H}^{2}$, i.e. a $S^{1}-$bundle
over $\mathbb{H}^{2}$ (or the tangent bundle where every vector has
norm one). The restricted bundle over $V^{2}$ is trivial so that
$UV^{2}=V^{2}\times S^{1}$. Let $L,L'$ be circular arcs (geodesics)
in $\mathbb{H}^{2}$ (invariant w.r.t. $SL(2,\mathbb{R})$) starting
at a common point which define parallel tangent vectors w.r.t. the
metrics of the upper half plane model. The foliation of $V^{2}$ is
given by geodesics transverse to the boundary and we obtain a foliation
of $V^{2}\times S^{1}$ (as unit tangent bundle). This foliation is
given by a $SL(2,\mathbb{R})$-invariant smooth 1-form $\omega$ (so
that $\omega=const.$defines the leaves) which is integrable $d\omega\wedge\omega=0$. 
\item With the relation $D^{2}=V^{2}\cup D_{1}^{2}\cup\cdots\cup D_{k-1}^{2}$,
we obtain $D^{2}\times S^{1}=V^{2}\times S^{1}\cup\left(D_{1}^{2}\times S^{1}\right)\cup\cdots\cup\left(D_{k-1}^{2}\times S^{1}\right)$
or the gluing of $k-1$ solid tori to $V^{2}\times S^{1}$ gives a
solid tori. Every glued solid torus will be foliated by a Reeb foliation.
Finally using $S^{3}=(D^{2}\times S^{1})\cup(S^{1}\times D^{2})$
(the Heegard decomposition of the 3-sphere) again with a solid torus
with Reeb foliation, we obtain a foliation on the 3-sphere. 
\end{enumerate}
The construction above will be also work for any 3-manifold. But now
we will define the Godbillon-Vey invariant or Godbillon-Vey class
for a 3-manifold. Let $N$ be a 3-manifold with a codimension-one
foliation $\mathcal{F}$. Then one has a smooth 1-form $\omega$ fulfilling
\[
d\omega\wedge\omega=0
\]
but defining another 1-form $\theta$ as the solution of the equation
\begin{equation}
d\omega=-\theta\wedge\omega\label{eq:def-equation-GV}
\end{equation}
so that 
\begin{equation}
\Gamma_{\mathcal{F}}=\theta\wedge d\theta\label{eq:Godbillon-Vey-class}
\end{equation}
is a closed 3-form. As discovered by Godbillon and Vey \cite{GodVey:71},
$\Gamma_{\mathcal{F}}$ depends only on the foliation $\mathcal{F}$
and not on the realization via $\omega,\theta$. Thus $\Gamma_{\mathcal{F}}$,
the \emph{Godbillon-Vey class}, is an invariant of the foliation.
The integral 
\[
GV(N,\mathcal{F})=\intop_{N}\theta\wedge d\theta
\]
known as Godbillon-Vey number has the properties: 
\begin{enumerate}
\item Let $\mathcal{F}_{1}$ and $\mathcal{F}_{2}$ be two cobordant foliations
then $\Gamma_{\mathcal{F}_{1}}=\Gamma_{\mathcal{F}_{2}}$ and $GV(N,\mathcal{F}_{1})=GV(N,\mathcal{F}_{2})$. 
\item The class $\Gamma$ vanishes for a Reeb foliation and $GV(N,\mathcal{F}_{Reeb})=0$. 
\item A decomposition $N=N_{1}\cup N_{2}$ respecting the foliation has
the Godbillon-Vey numbers $GV(N,\mathcal{F})=GV(N_{1},\mathcal{F}|_{N_{1}})+GV(N_{2},\mathcal{F}|_{N_{2}})$. 
\end{enumerate}
The Godbillon-Vey number was calculated for the foliation $\mathcal{F}_{Thurston}$
of the 3-sphere $S^{3}$. Using the decomposition 
\begin{equation}
S^{3}=\left(V^{2}\times S^{1}\right)\cup\left(D_{1}^{2}\times S^{1}\right)\cup\cdots\cup\left(D_{k-1}^{2}\times S^{1}\right)\cup\left(S^{1}\times D_{k}^{2}\right)\label{eq:decomposition-S3}
\end{equation}
with Reeb foliations $\mathcal{F}_{Reeb}$ for $D_{i}^{2}\times S^{1}$
and with the $SL(2,\mathbb{R})$-invariant foliation $\mathcal{F}_{SL}$
for $V^{2}\times S^{1}$, we get 
\begin{align*}
GV(S^{3},\mathcal{F}_{Thurston}) & =GV(V^{2}\times S^{1},\mathcal{F}_{SL})+\sum_{i=1}^{k}GV(D_{i}^{2}\times S^{1},\mathcal{F}_{Reeb})\\
 & =GV(V^{2}\times S^{1},\mathcal{F}_{SL})
\end{align*}
(using the properties of the Godbillon-Vey number above). Thurston
\cite{Thu:72} obtains for the Godbillon-Vey number 
\[
GV(V^{2}\times S^{1},\mathcal{F}_{SL})=4\pi\cdot vol(P)=8\pi\cdot vol(K)
\]
and 
\begin{equation}
GV(S^{3},\mathcal{F}_{Thurston})=4\pi\cdot Area(P)\label{eq:GV-number-Thurston-foliation-1}
\end{equation}
so that \emph{any real number can be realized by a suitable foliation
of this type}. Furthermore, two cobordant foliations have the same
Godbillon-Vey number (but the reverse is in general wrong). Let $[1]\in H^{3}(S^{3},\mathbb{R})$
be the dual of the fundamental class $[S^{3}]$ defined by the volume
form, then the Godbillon-Vey class can be represented by 
\begin{equation}
\Gamma_{\mathcal{F}_{a}}=4\pi\cdot Area(P)[1]\label{eq:Godbillon-Vey-class-Thurston-foliation}
\end{equation}
The Godbillon-Vey class is an element of the deRham cohomology $H^{3}(S^{3},\mathbb{R})$
which will be used later to construct a relation to gerbes. Furthermore
we remark that the classification is not complete, i.e. the kernel
of the map $GV$ is unknown. Thurston constructed only a surjective
homomorphism from the group of cobordism classes of foliation of $S^{3}$
into the real numbers $\mathbb{R}$. We remark the close connection
between the Godbillon-Vey class (\ref{eq:Godbillon-Vey-class}) and
the Chern-Simons form if $\theta$ can be interpreted as connection
of a suitable complex line bundle.

\subsection{Codimension-one foliations on 3-manifolds\label{sub:Codimension-one-foliations-on.3MF}}

Now we will discuss the general case of a compact 3-manifold carrying
a foliation of the same type like the 3-sphere above. The main idea
of the construction is very simple and uses a general representation
of all compact 3-manifolds by Dehn surgery. Here we will use an alternative
representation of surgery by using the Dehn-Lickorish theorem (\cite{PrasSoss:97}
Corollary 12.4 at page 84). Let $\Sigma$ be a compact 3-manifold
without boundary. There is now a natural number $k\in\mathbb{N}$
so that any orientable 3-manifold can be obtained by cutting out $k$
solid tori from the 3-sphere $S^{3}$ and then pasting them back in,
but along different diffeomorphisms of their boundaries. Moreover,
it can be assumed that all these solid tori in $S^{3}$ are unknotted.
Then any 3-manifold $\Sigma$ can be written as 
\[
\Sigma=\left(S^{3}\setminus\left(\bigsqcup_{i=1}^{k}D_{i}^{2}\times S^{1}\right)\right)\cup_{\phi_{1}}\left(D_{1}^{2}\times S^{1}\right)\cup_{\phi_{2}}\cdots\cup_{\phi_{k}}\left(D_{k}^{2}\times S^{1}\right)
\]
where $\phi_{i}:\partial\left(S^{3}\setminus\left(\bigsqcup_{i=1}^{k}D_{i}^{2}\times S^{1}\right)\right)\to\partial D_{i}^{2}\times S^{1}$
is the gluing map from each boundary component of $\left(S^{3}\setminus\left(\bigsqcup_{i=1}^{k}D_{i}^{2}\times S^{1}\right)\right)$
to the boundary of $\partial D_{i}^{2}\times S^{1}$. This gluing
map is a diffeomorphism of tori $T^{2}\to T^{2}$ (where $T^{2}=S^{1}\times S^{1}$).
The Dehn-Lickorish theorem describes all diffeomorphisms of a surface:
Every diffeomorphism of a surface is the composition of Dehn twists
and coordinate transformations (or small diffeomorphisms). The decomposition
(\ref{eq:decomposition-S3}) of the 3-sphere can be used to get a
decomposition of $\Sigma$ by 
\[
\Sigma=\left(V^{2}\times S^{1}\right)\cup_{\phi_{1}}\left(D_{1}^{2}\times S^{1}\right)\cup_{\phi_{2}}\cdots\cup_{\phi_{k}}\left(D_{k}^{2}\times S^{1}\right)
\]
which will guide us to the construction of a foliation on $\Sigma$: 
\begin{itemize}
\item Construct a foliation $\mathcal{F}_{\Sigma,SL}$ on $V^{2}\times S^{1}$
using a polygon $P$ (see above) and 
\item Glue in $k$ Reeb foliations of the solid tori using the diffeomorphisms
$\phi_{i}$. 
\end{itemize}
Finally we get a foliation $\mathcal{F}_{\Sigma,Thurston}$ on $\Sigma$.
According to the rules above, we are able to calculate the Godbillon-Vey
number 
\[
GV(\Sigma,\mathcal{F}_{\Sigma,Thurston})=4\pi\cdot vol(P)
\]
Therefore for any foliation of $S^{3}$, we can construct a foliation
on any compact 3-manifold $\Sigma$ with the same Godbillon-Vey number.
Both foliations $\mathcal{F}_{Thurston}$ and $\mathcal{F}_{\Sigma,Thurston}$
agree for the common submanifold $V^{2}\times S^{1}$ or there is
a foliated cobordism between $V^{2}\times S^{1}\subset\Sigma$ and
$V^{2}\times S^{1}\subset S^{3}$. Of course, $S^{3}$ and $\Sigma$
differ by the gluing of the solid tori but every solid torus carries
a Reeb foliation which does not contribute to the Godbillon-Vey number.
This claim completes the proof of the following Theorem: \begin{theorem}
\label{thm:foliation-3MF}Let $\Sigma$ be a compact 3-manifold without
boundary. Every codimension-one foliation $\mathcal{F}_{Thurston}$
of the 3-sphere $S^{3}$ induces a codimension-one foliation $\mathcal{F}_{\Sigma,Thurston}$
on $\Sigma$ so that the Godbillon-Vey numbers agree $GV(S^{3},\mathcal{F}_{Thurston})=GV(\Sigma,\mathcal{F}_{\Sigma,Thurston})$.
Furthermore there are diffeomorphic submanifolds $N\subset\Sigma$
and $N\subset S^{3}$ so that there is a foliated cobordism between
these two submanifolds with $GV(N,\mathcal{F}_{N,Thurston})=GV(S^{3},\mathcal{F}_{Thurston})$.
\end{theorem}

\subsection{Isotopy of surfaces and Rooted trees\label{sub:Isotopy-of-surfaces}}

This subsection is of technical nature. Here we will describe the
isotopy classes of embedded surfaces. Let $F$ be a surface together
with an embedding $F\hookrightarrow M$ in a manifold $M$. Two embedding
are isotopic if there is a homotopy (1-parameter family of maps) between
these two embeddings which are embeddings for all parameter values.
In our case, embeddings $F\hookrightarrow M$ are isotopic if there
is a diffeomorphism of the manifold $M$ and of the surface $F$ so
that the two embeddings agree. It is also possible to fix the manifold
$M$ and we have to consider the diffeomorphisms of $F$. According
to the Dehn-Lickorish theorem, every diffeomorphism of a surface splits
into a small diffeomorphism (a diffeomorphism connected to the identity
or a coordinate transformation in physics language) and a large diffeomorphism
(an element of the mapping class group). Large diffeomorphisms are
given by compositions of Dehn twists along a system of closed curves
(the \emph{essential curves}). For surfaces with boundary, one considers
a diffeomorphism which fixes the boundary pointwise. The product of
Dehn twists along a system of essential curves which intersect at
most ones are called tree-like mapping classes \cite{GerberTreeLikeMappingClasses2006}.
These classes can be also represented by a tree: every essential curve
is a vertex and the vertices of two, intersecting essential curves
are connected by an edge. Now it is possible to extract informations
about diffeomorphisms from the combinatorics of the tree. An important
fact is the relation to pseudo-Anosov diffeomorphisms for a large
class of trees \cite{ACampo1998}. Now we will specialize this theory
to a Casson handle. Casson handles will be introduced later, here
we need to know only that it is system of disks with self-intersection
arranged along a rooted tree. Every vertex of the tree is a set of
disks with self-intersections (the so-called tower) and every disk
with self-intersections is the beginning point of another tower (so
both vertices are connected by an edge). This construction will be
modified a little bit: one introduces towers of surfaces with non-zero
genus (called gropes) between the towers of self-intersecting disks.
The details of the construction will be explained later but for now
we have to consider the towers of self-intersecting disks and of non-zero
genus surfaces. Every tower of self-intersecting disks contains as
much essential disks as self-intersections (which will be killed by
the next tower to keep the Casson handle simply connected). Similarly,
the tower of surfaces admits also a system of essential curves (as
described by Lickorish in \cite{Lic:62}). Now a disk with one self-intersection
is not isotopic to a disk with two self-intersections (because both
fundamental groups are different and therefore there is no homotopy
between them). The same argument can be applied to the surfaces (but
additionally two surfaces with different genus are not homeomorphic).
Putting all parts together we can simply state: \emph{The isotopy
type of the Casson handle (and also of the capped grope) is determined
by the defining tree, i.e. two Casson handles with different trees
are non-isotopic to each other.} But we remark that we do not know
much about the smoothness structures of Casson handles (there are
uncountably many \cite{Gom:89} and countably many of them are determined
by the minimal intersection number \cite{Gom:84}).

\section{Small exotic $\mathbb{R}^{4}$ and codimension-one foliations\label{sec:Exotic-R4-codim-1-foliation}}

In this section we will describe the construction of small exotic
$\mathbb{R}^{4}$'s by using the failure of the smooth h cobordism
theorem for some compact 4-manifolds. The structure theorem for smooth
h cobordisms of compact 4-manifolds \cite{CuFrHsSt:97} singles the
Akbulut cork out as the reason for this failure. This description
has the great advantage to have an explicit coordinate representation.
The main part of this representation is given by an infinite construction
typically for 4-manifolds, the Casson handle. Here we can only present
the tip of the iceberg. There are very good books like the classical
ones \cite{Kir:89,FreQui:90} or modern views like \cite{GomSti:1999}
as well the original research articles \cite{Cas:73,Fre:79,Fre:82,Qui:82,Fre:83}
to understand Casson handles, capped gropes and its design. Using
this machinery, DeMichelis and Freedman \cite{DeMichFreedman1992}
were able to construct a family of uncountable many, non-diffeomorphic
exotic $\mathbb{R}^{4}$ (a radial family). Finally in subsection
\ref{sub:Exotic-R4-codim-1-foliation}, we use this radial family
to get a direct relation (Theorem \ref{thm:codim-1-foli-radial-fam})
to non-cobordant, codimension-one foliations of the 3-manifold $Y_{r}$
surrounding the member $R_{t}^{4}$ (with $t=1-1/r$) in the radial
family.

\subsection{The radial family of small exotic $\mathbb{R}^{4}$}

Let $\mathbf{R}^{4}$ be the standard $\mathbb{R}^{4}$ (i.e. $\mathbf{R}^{4}=\mathbb{R}^{3}\times\mathbb{R}$
smoothly) and let $R^{4}$ be a small exotic $\mathbb{R}^{4}$ (constructed
from the failure of the smooth h-cobordism, see \cite{Don:87}). Fix
a homeomorphism 
\[
h:\mathbf{R}^{4}\to R^{4}
\]
and define the image 
\[
R_{r}^{4}=h(B_{r}^{4})
\]
of an open ball $B_{r}^{4}=\left\{ x\in\mathbf{R}^{4}|\,||x||^{2}<r\right\} \subset\mathbf{R}^{4}$
of radius $r\in[0,\infty]$ centered at $0$ in $\mathbf{R}^{4}$.
We call $\left\{ R_{r}^{4}\right\} $ the \emph{radial family} with
$R_{\infty}^{4}=R^{4}$ (or $R_{\infty}^{4}=\mathbf{R}^{4}$ the standard
$\mathbb{R}^{4}$). For the following argumentation, we need the following
facts about the radial family (see chapter 9 in \cite{GomSti:1999}
and Fig. \ref{fig:Radial-family-exoticR4}): 
\begin{enumerate}
\item There is a compact submanifold $K\subset R^{4}$ which cannot be surrounded
by neighborhood with boundary a smoothly embedded 3-sphere. 
\item Every homeomorphism between 4-manifolds is isotopic to a local diffeomorphism
in the neighborhood of an preassigned 1-complex \cite{FreQui:90}.
Therefore, there is a topological radius function (polar coordinate)
$\rho:R^{4}\to[0,+\infty)$ so that $R_{t}^{4}=\rho^{-1}([0,r))$
with $t=1-1/r$. Let $\left\{ R_{t}^{4}\right\} $ with $t=1-1/r$
be this reparametrization of the radial family (with $t\in[0,1]$)
then $K\subset R_{1}^{4}$ and $K\subset R_{0}^{4}$ (see \cite{DeMichFreedman1992}
Theorem 3.2). 
\item For $s<t$ one has $R_{s}^{4}\subset R_{t}^{4}$ (the topological
radius function is continuous preserving therefore the order of the
interval, see the introduction of \cite{DeMichFreedman1992}). 
\item Every member of the radial family is an exotic $\mathbb{R}^{4}$ by
the defining property $K\subset R_{0}^{4}\subset R_{t}^{4}$ for all
$t\in(0,1]$. But there is a continuum of parameter values (as subset
of the Cantor set $CS$) so that the corresponding members $R_{s}^{4},R_{t}^{4}$
are pairwise non-diffeomorphic for $s\not=t\in CS$. 
\end{enumerate}
Every member of the radial family $R_{t}^{4}$ for $t\in[0,1]$ has
the property to contain a compact submanifold $K$ which cannot be
surrounded by a smooth 3-sphere. Therefore every member is an exotic
$\mathbb{R}^{4}$. Only for a Cantor subset $CS\subset[0,1]$, one
has an explicit description (using Casson handles and/or capped gropes)
and the members of this parameter set are non-diffeomorphic to each
other (see the introduction as well section 3 of \cite{DeMichFreedman1992}).
But where does the labeling with a Cantor set $CS$ came from? In
\cite{Fre:82}, Freedman constructed from a given Casson handle $Q$
(represented by a tree) a continuum of Casson handles which embed
into $Q$, called the design. This continuum is labeled by a Cantor
set and represented by a dyadic tree (see theorem 5.2 and 5.3 in \cite{Fre:82}).
But between these Casson handles (represented as paths in the dyadic
tree), there are other Casson handles, known as gaps. In the proof
of Theorem 1.1 in \cite{Fre:82}, one shrinks these gaps to get the
homeomorphism. The gaps are other Casson handles (and we have no control
about size and complexity). But we know that there is some tree representing
these handles. But (see subsection ?), at least all these Casson handles
are non-isotopic to each other (the Casson handles differ at least
in one self-intersection number). In \cite{DeMichFreedman1992}, one
can also find an explicit description of the members in the radial
family which was further investigated in \cite{Bizaca1995,BizGom:96}
for one example. We will come back to this example later. According
to these results, every member of the radial family has the common
compact subset $K$, the so-called Akbulut cork of the corresponding
h-cobordism, and a variable part given by a Casson handle or capped
grope (see below). Therefore, \emph{two members of the radial family
are at least non-isotopic to each other}.

\subsection{Casson handles}

Casson handles are important to understand 4-manifolds and especially
small exotic $\mathbb{R}^{4}$. Furthermore we will use a mixture
of Casson handle and capped grope techniques to proof our main theorem.
In particular it is easier to consider Casson handles (as an infinite
construction) first.

Let us now consider the basic construction of the Casson handle $CH$.
Let $M$ be a smooth, compact, simply-connected 4-manifold and $f:D^{2}\to M$
a (codimension-2) mapping. By using diffeomorphisms of $D^{2}$ and
$M$, one can deform the mapping $f$ to get an immersion (i.e. injective
differential) generically with only double points (i.e. $\#|f^{-1}(f(x))|=2$)
as singularities \cite{GolGui:73}. But to incorporate the generic
location of the disk, one is rather interesting in the mapping of
a 2-handle $D^{2}\times D^{2}$ induced by $f\times id:D^{2}\times D^{2}\to M$
from $f$. Then every double point (or self-intersection) of $f(D^{2})$
leads to self-plumbings of the 2-handle $D^{2}\times D^{2}$. A self-plumbing
is an identification of $D_{0}^{2}\times D^{2}$ with $D_{1}^{2}\times D^{2}$
where $D_{0}^{2},D_{1}^{2}\subset D^{2}$ are disjoint sub-disks of
the first factor disk. In complex coordinates the plumbing may be
written as $(z,w)\mapsto(w,z)$ or $(z,w)\mapsto(\bar{w},\bar{z})$
creating either a positive or negative (respectively) double point
on the disk $D^{2}\times0$. Consider the pair $(D^{2}\times D^{2},\partial D^{2}\times D^{2})$
and produce finitely many self-plumbings away from the attaching region
$\partial D^{2}\times D^{2}$ to get a kinky handle $(k,\partial^{-}k)$
where $\partial^{-}k$ denotes the attaching region of the kinky handle.
A kinky handle $(k,\partial^{-}k)$ is a one-stage tower $(T_{1},\partial^{-}T_{1})$
and an $(n+1)$-stage tower $(T_{n+1},\partial^{-}T_{n+1})$ is an
$n$-stage tower union of kinky handles $\bigcup_{\ell=1}^{n}(T_{\ell},\partial^{-}T_{\ell})$
where two towers are attached along $\partial^{-}T_{\ell}$. Let $T_{n}^{-}$
be $(\mbox{interior}T_{n})\cup\partial^{-}T_{n}$ and the Casson handle
\[
CH=\bigcup_{\ell=0}T_{\ell}^{-}
\]
is the union of towers (with direct limit topology induced from the
inclusions $T_{n}\hookrightarrow T_{n+1}$). A Casson handle is specified
up to (orientation preserving) diffeomorphism (of pairs) by a labeled
finitely-branching tree with base-point {*}, having all edge paths
infinitely extendable away from {*}. Each edge should be given a label
$+$ or $-$ and each vertex corresponds to a kinky handle; the self-plumbing
number of that kinky handle equals the number of branches leaving
the vertex. The sign on each branch corresponds to the sign of the
associated self plumbing. The whole process generates a tree with
infinite many levels. In principle, every tree with a finite number
of branches per level realizes a corresponding Casson handle. The
simplest non-trivial Casson handle is represented by the tree $Tree_{+}$:
each level has one branching point with positive sign $+$. The reverse
construction of a Casson handle $CH_{\mathcal{T}}$ by using a labeled
tree $\mathcal{T}$ can be found in the appendix A. Let $\mathcal{T}_{1}$
and $\mathcal{T}_{2}$ be two trees with $\mathcal{T}_{1}\subset\mathcal{T}_{2}$
(it is subtree) then $CH_{\mathcal{T}_{2}}\subset CH_{\mathcal{T}_{1}}$.

\subsection{Capped gropes and its design\label{sub:Capped-gropes}}

This subsection is a very technical approach using the design of Freedman.
The main idea is a foliation of a Casson handle by using the frontier
(a kind of boundary) for all sub-Casson handles. We need this part
of the theory to get a foliation for each member of the radial family.

The modern way to the classification of 4-manifolds used ``capped
gropes'', a mixed variant of Casson handle and grope (chapters 1
to 4 in \cite{FreQui:90}). These differ from Casson handles in that
many surface stages are interspersed between the immersed disks of
Casson's construction. The growth rate of their stages was determined
in \cite{AncelStarbird1989} (Theorem A) to be at least exponential
(more than $2^{n}$).

\begin{figure}
\centerline{\psfig{file=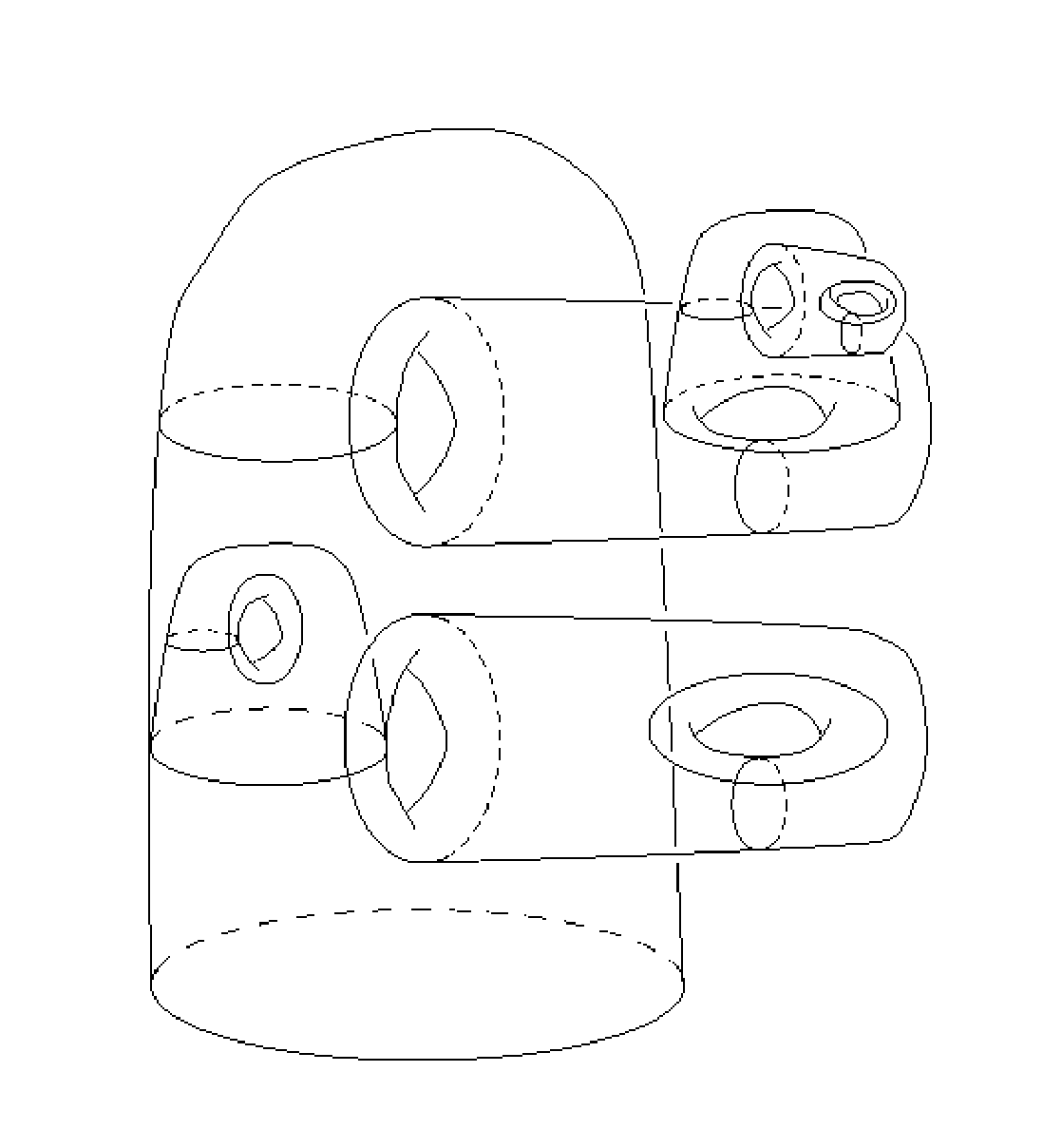,width=6cm}}
\caption{Example of a grope with symplectic basis as curves around the holes\label{fig:Example-of-grope}}
\end{figure}

A grope is a special pair (2-complex,circle), where the circle is
referred to as the boundary of the grope. There is an anomalous case
when the depth is $1$: the unique grope of depth 1 is the pair (circle,circle).
A grope of depth 2 is a punctured surface with the boundary circle
specified (see Fig. \ref{fig:Example-of-grope}). 

To form a grope $G$ of depth $n$, take a punctured surface, $F$,
and prescribe a symplectic basis $\left\{ \alpha_{i},\beta_{j}\right\} $.
That is, $\alpha_{i}$ and $\beta_{j}$ are embedded curves in $F$
which represent a basis of $H_{1}(F)$ such that the only intersections
among the $\alpha_{i}$ and $\beta_{j}$ occur when $\alpha_{i}$
and $\beta_{j}$ meet in a single point $\alpha_{i}\cdot\beta_{j}=1$.
Now glue gropes of depth $<n$ along their boundary circles to each
$\alpha_{i}$ and $\beta_{j}$ with at least one such added grope
being of depth $n-1$. (Note that we are allowing any added grope
to be of depth 1, in which case we are not really adding a grope.)
The surface $F\subset G$ is called the bottom stage of the grope
and its boundary is the boundary of the grope. The tips of the grope
are those symplectic basis elements of the various punctured surfaces
of the grope which do not have gropes of depth $>1$ attached to them.
\emph{A capped grope is a grope with disks (the caps) attached to
all its tips.} The grope without caps is sometimes called the body
of the capped grope. The capped grope (as cope) was firstly described
by Freedman in 1983 \cite{Fre:83}. A capped grope is built of layers
of surfaces, and a layer of disks (the caps) at the top. The caps
are only immersed disks like in case of the Casson handle to make
the grope simply-connected. In chapter 3 of \cite{FreQui:90}, a tower
construction was defined whose building blocks are themselves capped
gropes. Especially the convergence of infinite towers is discussed.
We refer to this chapter for the details but remark, that one replace
the immersed disks (the caps) by Casson handles. Therefore the Casson
handle is a special case of a capped grope (surface of genus $0$
with one cap). The (convergent) towers of capped grope (or capped
towers) can be parametrized (by a singular parametrization, see subsection
4.2 in \cite{FreQui:90}). The detailed description of this parametrization
is given in subsection 4.3 in \cite{FreQui:90} by using the middle
third's Cantor set. For simplicity we will introduce the parametrization
of a Casson handle, the case of a capped grope is similar. Especially
the argumentation using sequences of $0$'s and $2$'s is the same
for Casson handles and capped gropes.

A Casson handle is represented by a labeled finitely-branching tree
$Q$ with base point $\star$, having all edge paths infinitely extendable
away from $\star$. Freedman (\cite{Fre:82} p.398) constructed another
labeled tree $S(Q)$ from the tree $Q$. There is a base point from
which a single edge (called ``decimal point'') emerges. The tree
is binary: one edge enters and two edges leaving a vertex. The edges
are named by initial segments of infinite base 3-decimals representing
numbers in the standard middle third Cantor set $CS\subset[0,1]$.
This kind of Cantor set is given by the following construction: Start
with the unit interval $S_{0}=[0,1]$ and remove from this set the
middle third and set $S_{1}=S_{0}\setminus(1/3,2/3)$. Continue in
this fashion, where $S_{n+1}=S_{n}\setminus\{\mbox{middle thirds of subintervals of }S_{n}\}$.
Then the Cantor set $CS$ is defined as $CS=\cap_{n}S_{n}$. With
other words, if we using a ternary system (a number system with base
3), then we can write the Cantor set as $CS=\left\{ x=(0.a_{1}a_{2}a_{3}\ldots)\mbox{ where each }a_{i}=0\mbox{ or }2\right\} $.
Each edge $e$ of $S(Q)$ carries a label $\tau_{e}$ where $\tau_{e}$
is an ordered finite disjoint union of 5-stage towers (see the previous
subsection) together with an ordered collection of standard loops
generating the fundamental group. There are three constraints on the
labels which lead to the correspondence between the $\pm$ labeled
tree $Q$ and the (associated) $\tau$-labeled tree $S(Q)$. One calls
$S(Q)$ the design.

Two words are in order for the design $S(Q)$: first, every sequence
of $0$'s and $2$'s is one path in $S(Q)$ representing one embedded
capped gropes $GCH_{Q_{1}}\subset GCH_{Q}$ where both trees are related
like $Q\subset Q_{1}$. For example, the Casson handle corresponding
to $.020202...$ is obtained as the union of the 5-stage towers $T^{0}\cup T^{02}\cup T^{020}\cup T^{0202}\cup T^{02020}\cup T^{020202}\cup...$.
For later usage we identify the sequence $.00000...$ with the Tree
$Tree_{+}$(normalization). Secondly, there are gaps, i.e. we have
only a Cantor set of Casson handles. For instance a gap is lying between
the paths $.022222\ldots$ and $.20000\ldots$ In the proof of Freedman,
a neighborhood of the gaps is shrunk to a point and one gets the desired
homeomorphism. But the gaps are also important to understand the design
as well its frontier. The frontier of a set $K$ is defined by $Fr(K)=\mbox{closure}(\mbox{closure}(K)\setminus K)$.
As example we consider the interior $int(D^{2})$ of a disk and obtain
for the frontier $Fr(int(D^{2}))=\mbox{closure}(\mbox{closure}(int(D^{2}))\setminus int(D^{2}))=\partial D^{2}$,
i.e. the boundary of the disk $D^{2}$. The frontier and gaps are
the big difference between a Casson handle and a capped grope. The
frontier of the Casson handle is the manifold factor $S^{1}\times D^{2}/Wh$
(that is, it becomes manifold upon crossing with the real line), but
not a manifold ($Wh$ is the Whitehead continuum, see \cite{Whitehead35,Fre:82}).
In contrast, the frontier of the capped grope is the solid torus $S^{1}\times D^{2}$
(see the shrinking argument in \cite{AncelStarbird1989}). The gaps
are very complicated in case of a Casson handle, it looks like $S^{1}\times D^{3}/Wh$
whereas for the capped grope it is $S^{1}\times D^{3}$. In the following
we will explain the common structure of the parametrization or design
beginning with the Casson handle but then switching to the (convergent)
capped tower. The difference is only given by the simpler structure
of the gaps ($S^{1}\times D^{3}$) and the frontier ($S^{1}\times D^{2}$).

Let $\gamma\in S(Q)$ be a path and $F_{\gamma}$ the frontier with
its closure $S^{1}\times D^{2}/Wh_{\gamma}$ (see \cite{Fre:82}).
The interior of every Casson handle is diffeomorphic to the standard
$\mathbb{R}^{4}$. Therefore the frontier $F_{\gamma}$ is most important
to understand the Casson handle. Now consider the two paths $\gamma=.022222\ldots$
and $\gamma'=.20000\ldots$. Every path in $S(Q)$ is represented
by one sequence over the alphabet $\{\text{0,2\}}$. Every gap is
a sequence containing at least one $1$ (so for instance $.1222...$
or $.012222...)$. In fact, the gap between $\gamma$ and $\gamma'$
corresponds to the first middle third which is deleted by constructing
$S(Q)$ and one has the relation $\gamma\cup\gamma'=\partial[.1n_{1}n_{2}\ldots]=\partial[\mbox{all numbers beginning .1}]$
(see chapter XIII \S4 of \cite{Kir:89}). Therefore the boundaries
of the gaps are the important part to understand the design.

\begin{figure}
\centerline{\psfig{file=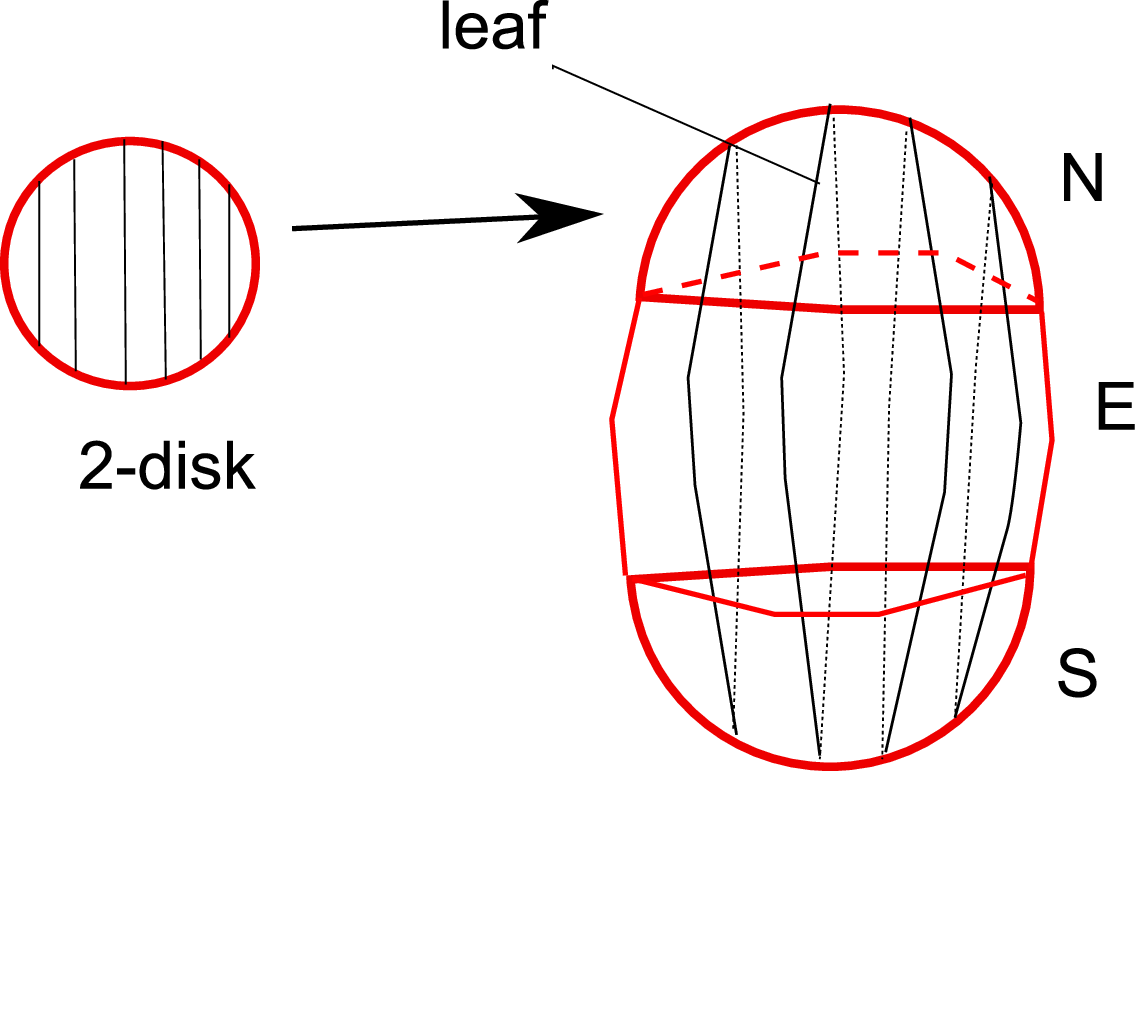,width=7cm}}
\caption{foliation of the 2-sphere as foliated cobordism $E$ of the two disks
$N,S$ \label{fig:foliation-of-S2-1}}
\end{figure}

Here we will use this structure to produce a foliation of the design.
There is now a natural order structure given by the sequence (for
instance $.022222...<.12222...<.22222...$). The leaves are the corresponding
gaps or Casson handles (represented by the union 5-stage towers ending
with $T^{02222...},\, T^{12222..}\mbox{ or }T^{22222...}$). The tree
structure of the design $S(Q)$ should be also reflected in the foliation
to represent every path in $S(Q)$ as a union of 5-stage towers. By
the reembedding theorems, the 5-stage towers can be embedded into
each other. Then we obtain two foliations of the (topological) open
2-handle $D^{2}\times\mathbb{R}^{2}$: a codimension-one foliation
along one $\mathbb{R}-$axis labeled by the sequences (for instance
$.022222...<.12222...<.22222...$) and a second codimension-one foliation
along the radius of the disk $D^{2}$ induced by inclusion of the
5-stage towers (for instance $T^{0}\supset T^{02}\supset T^{020}\supset...$).
The exploration of a Casson handle by using the design is given by
its frontier, in this case, minus the attaching region. In case of
a usual tower we get the frontier $S^{1}\times D^{2}/Wh_{\gamma}$
with $\gamma\in S(Q)$. The gaps have a similar structure. Then the
foliation of the Casson handle (induced from the design) is given
by the leaves $S^{1}\times D^{1}$ over the disk $D^{2}$ in the Casson
handle, i.e. the disk $D^{2}$ is foliated by parallel lines (see
Fig. \ref{fig:foliation-of-the-disk}). So, every Casson handle with
a given tree $Q$ has a codimension-one foliation given by its design.
This foliation can be also understood as a foliated cobordism. \begin{lemma}
\label{lem:foliation-gap-1}The foliation of the design is induced
by a foliation of the gaps seen as foliated cobordism and vice verse.
\end{lemma} \emph{Proof: }Gaps are characterized as sequences containing
at least a $1$, so that we have a countable number of gaps. Above
we discussed the relation 
\[
\gamma\cup\gamma'=\partial[.1n_{1}n_{2}\ldots]=\partial[\mbox{all numbers beginning .1}]
\]
between the two paths $\gamma=.022222\ldots$ and $\gamma'=.20000\ldots$
and the gap $.1n_{1}n_{2}\ldots$ . Gaps in the Casson handle look
like $S^{1}\times D^{3}/Wh$. Therefore we have to concentrate on
$S^{1}\times D^{3}$ and its boundary (or better its frontier) $S^{1}\times S^{2}$
(or better $S^{1}\times S^{2}/Wh$) to understand the foliation of
the whole design. For that purpose we consider the foliation as part
of a foliation of the 2-sphere (see Fig. \ref{fig:foliation-of-S2-1})
like in the example at the end of section \ref{sub:Definition-of-Foliation-cobordism}.

The 2-sphere is decomposed by $S^{2}=N\cup E\cup S$, two pole regions
$N,S$ ($N,S=D^{2})$ and an equator region $E=S^{1}\times D^{1}$.
The foliation of the disk as in Fig. \ref{fig:foliation-of-the-disk}
can be used to foliate $N$ and $S$. The foliation of the disk can
be connected by the leaves $S^{1}$ which are the latitudes. Then
one obtains a foliated cobordism $D^{3}$ (see the examples at the
end of subsection \ref{sub:Definition-of-Foliation-cobordism}) between
$N$ and $S$ given by the obvious foliation of the equator region
$E$ (a cylinder).$\square$

\subsection{Exotic $\mathbb{R}^{4}$ and codimension-one foliations\label{sub:Exotic-R4-codim-1-foliation}}

In this subsection we will construct a codimension-one foliation with
non-trivial Godbillon-Vey invariant on the 3-manifold $Y_{n}$ separating
the compact set $K$ (used in the construction of the small exotic
$\mathbb{R}^{4}$) from infinity. This foliation is directly related
to the exotic smoothness structure. This subsection is very technical
and the most complicated part of the paper.

\begin{figure}
\centerline{\psfig{file=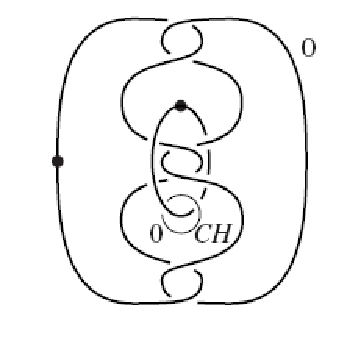,width=6cm}}
\caption{link picture for the compact subset $K$\label{fig:link-picture-for-K}}
\end{figure}

The strategy of the proof goes like this:

Let $R_{t}^{4}$ be a fixed radial family defined by a topological
radius function $\rho:R^{4}\to[0,\infty)$ so that $R_{t}^{4}=\rho^{-1}([0,r))$
with $t=1-1/r$. By definition, $R_{t}^{4}$ is embedded in any other
member $R_{u}^{4}$ for $t<u$. The completion $\bar{R}_{t}^{4}$
of $R_{t}^{4}\subset R_{u}^{4}$ is formally given by $\bar{R}_{t}^{4}=\rho^{-1}([0,r])$
with boundary the 3-manifold $Y_{r}$. There is also the possibility
to construct $Y_{r}$ directly as the limit $n\to\infty$ of a sequence
$\left\{ Y_{n}\right\} $ of 3-manifolds. To construct this sequence
of 3-manifolds, one can use the Kirby calculus, i.e. one represents
the compact subset $K$ by 1- and 2-handles pictured by a link say
$L_{K}$ where the 1-handles are represented by a dot (so that surgery
along this link gives $K$). Then one attaches a Casson handle to
this link. As an example see Fig. \ref{fig:link-picture-for-K}. 

\begin{figure}
\centerline{\psfig{file=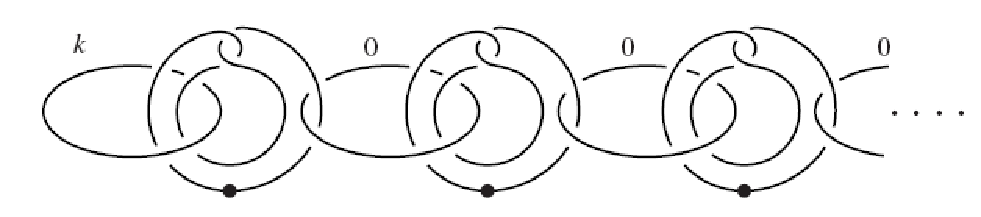,width=8cm}\quad{}\psfig{file=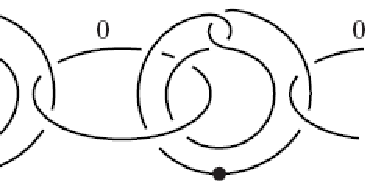,width=4cm}}
\caption{building block of every Casson handle (right) and the simplest Casson
handle (left)\label{fig:building-block-simplest-CH}}
\end{figure}

The Casson handle is given by a sequence of Whitehead links (where
the unknotted component has a dot) which are linked according to the
tree (see the right figure of Fig. \ref{fig:building-block-simplest-CH}
for the building block and the left figure for the simplest Casson
handle given by the unbranched tree). 

\begin{figure}
\centerline{\psfig{file=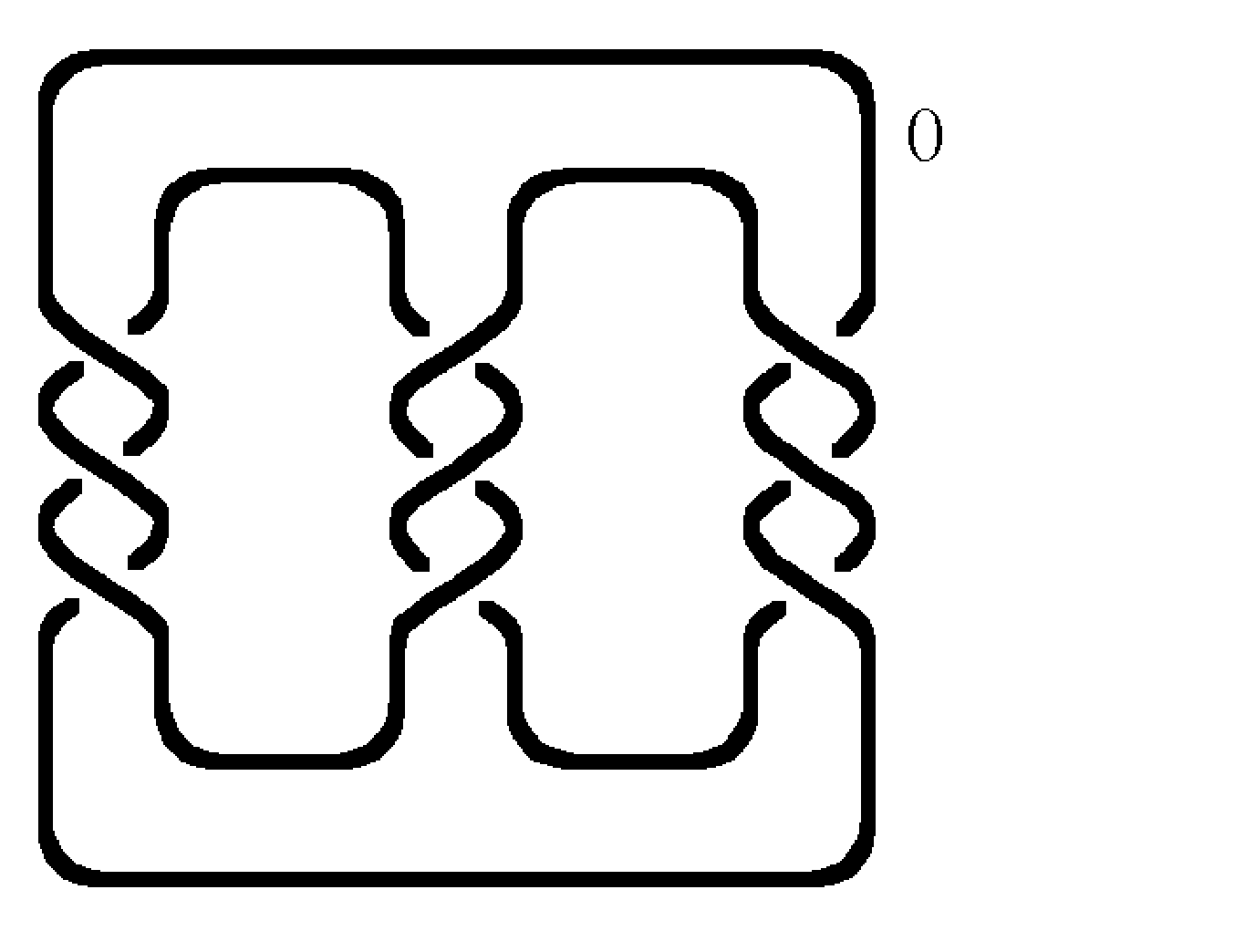,width=4cm}}
\caption{pretzel knot $(-3,3,-3)$ or the knot $9_{46}$ in Rolfson notation
producing the 3-manifold $Y_{1}$ by $0-$framed Dehn surgery\label{fig:pretzel-knot}}
\end{figure}

For the construction of a 3-manifold which surrounds the compact $K$,
one consider $n-$stages of the Casson handle and transforms the diagram
to a real link (the dotted components become usual components with
framing $0$). By a handle manipulations one obtains a knot so that
the $n$th (untwisted) Whitehead double of this knot represents the
desired 3-manifold (by using surgery). Then our example in Fig. \ref{fig:link-picture-for-K}
will result in the $n$th untwisted Whitehead double of the pretzel
knot $(-3,3,-3)$, Fig. \ref{fig:pretzel-knot} (see \cite{Ganzel2006}
for the handle manipulations). 

Then $0-$framed surgery along this pretzel knot produces $Y_{1}$
whereas the $n$th untwisted Whitehead double will give $Y_{n}$.
For large $n$, the structure of the Casson handle is contained in
the topology of $Y_{n}$ and in the limit $n\to\infty$ we obtain
$Y_{r}$ (which is now a wild embedding $Y_{r}\subset R_{u}^{4}$
in a larger member of the radial family, see above). But what did
we know about the structure of $Y_{n}$ or $Y_{r}$ in general? The
compact subset $K$ is topologically contractable having a topological
3-sphere as boundary. By construction using the failure of the h-cobordism
theorem \cite{CuFrHsSt:97}, $K$ is also a smoothly contractable
4-manifold having the boundary of a homology 3-sphere (see \cite{Fre:82}).
This information is also contained in $Y_{r}$. By the prime decomposition
\cite{Mil:62} (see also Appendix B), every homology 3-sphere is a
connected sum of irreducible 3-manifolds, Brieskorn homology 3-spheres
(Seifert fibered spaces) and hyperbolic homology 3-spheres where the
Poincare 3-sphere is ruled out (by Donaldsons theorem \cite{Don:83}).
But by the geometrization theorem (conjectured by Thurston \cite{Thu:97}
and proved by Perelman, see appendix B), there are two different geometries
for the irreducible pieces: it carries a hyperbolic geometry or the
so-called $\tilde{SL}_{2}$ geometry (see appendix B or \cite{Scott1983}
for a definition). But every $Y_{n}$ is given by an untwisted Whitehead
double of some knot. Using surgery, the geometry is determined by
the complement of the Whitehead double of the knot. But this complement
is a hyperbolic 3-manifold (see \cite{Budney2006}, the splicing of
the Whitehead link and the knot is the Whitehead double, see also
our example: the pretzel knot is hyperbolic). Now we need another
fundamental result: every diffeomorphism of a compact hyperbolic 3-manifold
(of finite volume) is an isometry (Mostow-Prasad rigidity) \cite{Mos:68}
or expressed differently, the volume of a compact, hyperbolic 3-manifold
is a topological invariant (see appendix B again). Therefore, the
\emph{compact 3-manifold $Y_{r}$ contains a 3-dimensional submanifold
$Y|_{\mathbb{H}}\subset Y_{r}$ with hyperbolic geometry having a
fixed size}. This size is of course relative to the embedding space,
i.e. relative to the radial family. But the foliation of the radial
family with respect to the radius (using the topological function)
implies (via the homeomorphism $h(B_{r}^{4})$ with scaling $vol(B_{r}^{4})\sim r^{3}$)
that the size of this hyperbolic 3-submanifold $Y|_{\mathbb{H}}$
must be scale like the radius $r$, i.e. 
\[
vol(Y|_{\mathbb{H}})\sim r^{3}
\]
whereas the size of the other components cannot be controlled (and
we scale it to be a submanifold of small size). Therefore we showed:
\begin{lemma} \label{lem:size-control-Y} Let $Y_{r}$ be the completion
of the member $R_{t}^{4}$ (with $t=1-1/r$). There is a submanifold
$Y|_{\mathbb{H}}\subset Y_{r}$ admitting a hyperbolic metric in the
interior. The volume of $Y|_{\mathbb{H}}$ is rigid and depends on
the radius $r$ like $vol(Y|_{\mathbb{H}})\sim r^{3}$.\end{lemma}
Using this geometrical input together with Gabai's work \cite{Gabai1983},
one has abstractly a codimension-one foliation for every component
of $Y_{r}$. The gluing of foliations is now the problem. But the
complexity of the Casson handle is encoded into the structure of the
gluing so that we will get a relation between singular foliations
and the tree representing the Casson handle. \begin{lemma} \label{lem:foliation-Y-for-member}
Let $Y_{r}$ be a compact 3-manifold as above. There is a codimension-one
foliation of $Y_{r}$ with Godbillon-Vey number $GV(Y_{r})\sim r^{2}$
constructed from the Casson handle for the corresponding member $R_{t}^{4}$.
\end{lemma} \emph{Proof:} As explained above, there is a hyperbolic
submanifold $Y|_{\mathbb{H}}\subset Y_{r}$ with scaling $vol(Y|_{\mathbb{H}})\sim r^{3}$.
By the Dehn-Lickorish theorem, we have a decomposition 
\[
Y|_{\mathbb{H}}=\left(V^{2}\times S^{1}\right)\cup_{\phi_{1}}\left(D_{1}^{2}\times S^{1}\right)\cup_{\phi_{2}}\cdots\cup_{\phi_{k}}\left(D_{k}^{2}\times S^{1}\right)
\]
in the notation of subsection \ref{sub:Codimension-one-foliations-on.3MF}.
The surface $V^{2}$ is a hyperbolic surface (it has negative Euler
characteristics) where the hyperbolic structure of the surface is
induced from the hyperbolic structure of the 3-manifold (using the
subgroup relation $PSL(2,\mathbb{R})\subset PSL(2,\mathbb{C})$ of
the isometry groups). Then $V^{2}\times S^{1}$ can be interpreted
as the unit tangent $UV^{2}$ of $V^{2}$. The isometry group of $V^{2}$
has to be identified with $PSL(2,\mathbb{R})$ (fixing the size of
$V^{2}$). The scaling behavior must be $vol(V^{2})\sim r^{2}$ by
the isometry group $PSL(2,\mathbb{R})$ and by the rigidity of $Y|_{\mathbb{H}}$
(otherwise it would contradict the rigidity of $Y|_{\mathbb{H}}$).
This behavior is further supported by Gabai's work \cite{Gabai1983}
that a hyperbolic 3-manifold admits a codimension-one foliation. According
to our argumentation above, the foliation of the unit tangent bundle
must be $PSL(2,\mathbb{R})$ invariant. It is a smooth codimension-one
foliation, i.e. there is a $PSL(2,\mathbb{R})$-invariant 1-form $\omega$
fulfilling $\omega\wedge d\omega=0$ (integrability). Using (\ref{eq:def-equation-GV}),
one has the Godbillon-Vey class (\ref{eq:Godbillon-Vey-class}). Finally
we obtain the Godbillon-Vey number for this foliation to be (\ref{eq:GV-number-Thurston-foliation-1})
and therefore 
\[
GV(Y_{r})\sim vol(V^{2})\sim r^{2}
\]
finishing the proof. $\square$

\begin{figure}
\centerline{\psfig{file=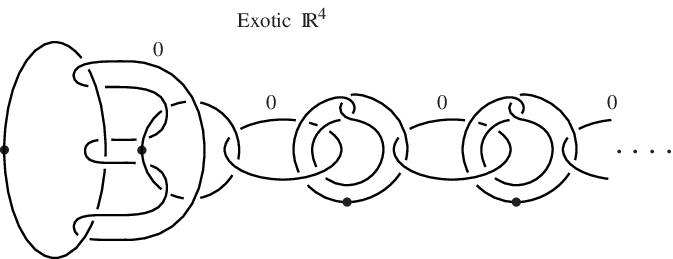,width=12cm}}
\caption{Handle picture of the small exotic $\mathbb{R}^{4}$, the components
with the dot are 1-handles and without the dot are 2-handles\label{fig:Handle-picture-of-exotic-R4}}
\end{figure}

This result is only partly satisfactory. The member of the radial
family $R_{t}^{4}$ is described by the 3-manifold $Y_{r}$ in its
full complexity. But the radius parameter $r$ (or $t$ with $t=1-1/r$)
is not a direct way to describe the complexity of the exotic $\mathbb{R}^{4}$
because it is not diffeomorphism invariant. We know that a small exotic
$\mathbb{R}^{4}$ can be described by a compact 4-manifold with an
attached Casson handle (see \cite{BizGom:96}). Therefore we will
give an intrinsic definition using only the complexity of the Casson
handle and of the compact 4-manifold. At first we will formulate our
result. \begin{theorem}\label{thm:foliation-single-small-exotic-R4}
Let $R^{4}$ be a small exotic $\mathbb{R}^{4}$ together with an
embedding $I:R^{4}\hookrightarrow S^{4}$ where $R^{4}$ is decomposed
like $R^{4}=K\cup CH_{\mathcal{T}}$ with the tree $\mathcal{T}$
. The embedding $I$ induces an embedding of the tree $\mathcal{T}\hookrightarrow D^{2}\subset\mathbb{H}^{2}$
with $vol(D^{2})\sim r^{2}$ (for some $r\in\mathbb{R}_{+}$). The
closure $\overline{I(R^{4})}$ has the boundary $Y_{\mathcal{T}}=\partial\left(\overline{I(R^{4})}\right)$.
Then $Y_{\mathcal{T}}$ admits a codimension-one foliation with Godbillon-Vey
number $GV\sim r^{2}$. \end{theorem} \emph{Proof: }We will demonstrate
it using the example of \cite{Biz:95,BizGom:96} which was partly
explained above. The general case is similar and we will explain the
differences. The interior of the handle body in Fig. \ref{fig:Handle-picture-of-exotic-R4}
is the $R^{4}$. 

\begin{figure}
\centerline{\psfig{file=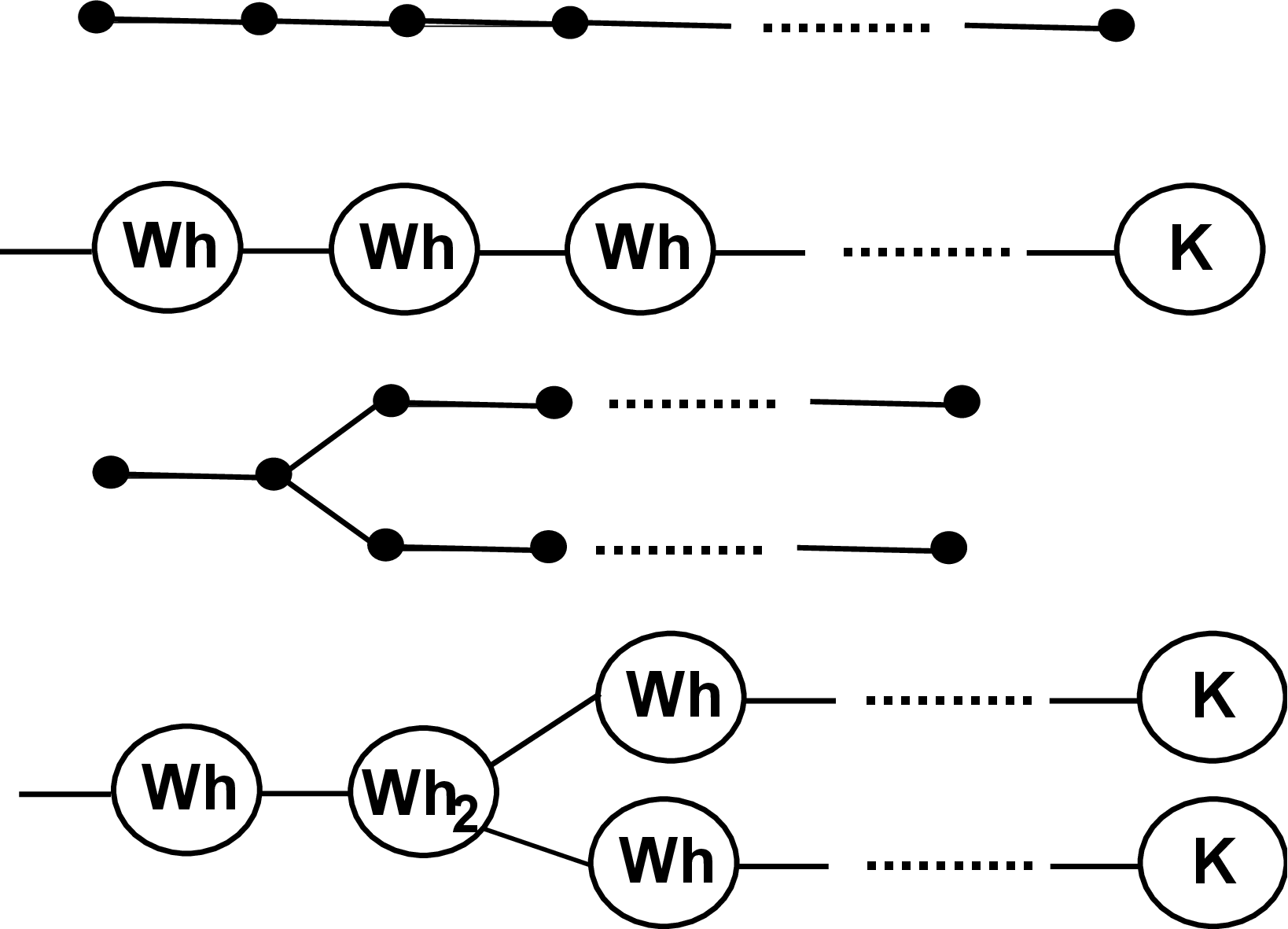,width=8cm}}
\caption{Schematic picture for the splitting of the knot complement $C(\mathcal{K}_{+})$
(above) and in the more general case $C(\mathcal{K_{T}})$ (below)\label{fig:Splitting-of-knot-complement} }
\end{figure}

The Casson handle is given by the simplest tree $\mathcal{T}_{+}$,
one positive self-intersection for each level. The compact 4-manifold
inside of $R^{4}$ can be seen in Fig. \ref{fig:link-picture-for-K}
as handle body. The 3-manifold $Y_{n}$ surrounding this compact submanifold
$K$ is given by surgery ($0-$framed) along the link in Fig. \ref{fig:Handle-picture-of-exotic-R4}
with a Casson handle of $n-$levels. In \cite{Ganzel2006}, this case
is explicitly discussed. $Y_{n}$ is given by $0-$framed surgery
along the $n$th untwisted Whitehead double of the pretzel $(-3,3-3)$
knot (see Fig. \ref{fig:pretzel-knot}). Obviously, there is a sequence
of inclusions 
\[
\ldots\subset Y_{n-1}\subset Y_{n}\subset Y_{n+1}\subset\ldots\to Y_{\mathcal{T}_{+}}
\]
with the 3-manifold $Y_{\mathcal{T}_{+}}$ as limit. Let $\mathcal{K}_{+}$
bet the corresponding (wild) knot, i.e. the $\infty$th untwisted
Whitehead double of the pretzel knot $(-3,3,-3)$ (or the knot $9_{46}$
in Rolfson notation). The surgery description of $Y_{\mathcal{T}_{+}}$induces
the decomposition 
\begin{equation}
Y_{\mathcal{T}_{+}}=C(\mathcal{K}_{+})\cup\left(D^{2}\times S^{1}\right)\qquad C(\mathcal{K}_{+})=S^{3}\setminus\left(\mathcal{K}_{+}\times D^{2}\right)\label{eq:surgery-description-of-Y}
\end{equation}
where $C(\mathcal{K}_{+})$ is the knot complement of $\mathcal{K}_{+}$.
In \cite{Budney2006}, the splitting of knot complements was described.
Let $K_{9_{46}}$ be the pretzel knot $(-3,3,-3)$ and let $L_{Wh}$
be the Whitehead link (with two components). Then the complement $C(K_{9_{46}})$
has one torus boundary whereas the complement $C(L_{Wh})$ has two
torus boundaries. Now according to \cite{Budney2006}, one obtains
the splitting 
\[
C(\mathcal{K}_{+})=C(L_{Wh})\cup_{T^{2}}C(L_{Wh})\cup_{T^{2}}\cdots\cup_{T^{2}}C(L_{Wh})\cup_{T^{2}}C(K_{9_{46}})
\]
and we will describe each part separately (see Fig. \ref{fig:Splitting-of-knot-complement}).

At first the knot $K_{9_{46}}$ is a hyperbolic knot, i.e. the interior
of the 3-manifold $C(K_{9_{46}})$ admits a hyperbolic metric. By
the work of Gabai \cite{Gabai1983}, $C(K_{9_{46}})$admits a codimension-one
foliation. The Whitehead link is a hyperbolic link but we need more:
the Whitehead link is a fibered link of genus $1$. That is, there
is a fibration of the link complement $\pi:C(L_{Wh})\to S^{1}$ over
the circle so that $\pi^{-1}(p)$ is a surface of genus $1$ (Seifert
surface) for all $p\in S^{1}$. The hyperbolic structure of $C(L_{Wh})$
with isometry group $PSL(2,\mathbb{C})$ induces by the subgroup relation
$PSL(2,\mathbb{R})\subset PSL(2,\mathbb{C})$ a hyperbolic structure
on the surface, i.e. the surface is given by a polygon with four sides
in $\mathbb{H}^{2}$. The foliation of this polygon is $PSL(2,\mathbb{R})$
by construction, i.e. it has $GV\not=0$. Then $C(L_{Wh})\cup_{T^{2}}C(L_{Wh})$
also fibers over $S^{1}$ and the polygon in $\mathbb{H}^{2}$ is
now the sum of the two polygons along one side. This procedure can
be done for all complements $C(L_{Wh})$ in $C(\mathcal{K}_{+})$
to get a polygon $P$ in $\mathbb{H}^{2}$. Now we can use the embedding
of the tree $\mathcal{T}_{+}\hookrightarrow D^{2}\subset\mathbb{H}^{2}$
to get an embedding of the polygon $P\hookrightarrow D^{2}$. The
Godbillon-Vey numbers are additive and we obtain by the embedding
$GV\sim vol(D^{2})\sim r^{2}$ as upper value. There is an additive
value of $GV$ for $C(K_{9_{46}})$ which does not change anything
for the scaling behavior of $GV\sim r^{2}$. The $D^{2}\times S^{1}$
component in (\ref{eq:surgery-description-of-Y}) can be chosen
to be a Reeb foliation with vanishing $GV$ number. It finish the
proof but we will also describe the changes for a general tree. At
first we will modify the Whitehead link: we duplicate the linked circle,
i.e. there are as many circles as branching in the tree to get the
link $Wh_{n}$ with $n+1$ components. Then the complement of $Wh_{n}$
has also $n+1$ torus boundaries and it also fibers over $S^{1}$.
With the help of $Wh_{n}$ we can build every tree $\mathcal{T}$.
Now the 3-manifold $Y_{\mathcal{T}}$ is given by $0-$framed surgery
along the $\infty$th untwisted ramified (usage of $Wh_{n}$) Whitehead
double of a knot $k$, denoted by the link $\mathcal{K_{T}}$. The
tree $\mathcal{T}$ has $m$ endpoints ($m$ can be also infinite),
then $Y_{\mathcal{T}}$ is given by 
\[
Y_{\mathcal{T}}=C(\mathcal{K_{T}})\cup\bigsqcup_{m}\left(D^{2}\times S^{1}\right)
\]
and the complement $C(\mathcal{K_{T}})$ splits like the tree into
complements of $Wh_{n}$ and $m$ copies of $C(k)$ (see Fig. \ref{fig:Splitting-of-knot-complement}).
All argumentation above can be extended to the general case and the
embedding of the tree $\mathcal{T}$ guarantees the finiteness of
the Godbillon-Vey number as well the scaling $GV\sim vol(D^{2})\sim r^{2}$.
$\square$

The results above showed two facts: after the embedding of the tree
(representing the Casson handle) one fixes the scale $r$ of the 3-manifold
$Y$ surrounding the small exotic $\mathbb{R}^{4}$ embedded in the
4-sphere (or in the radial family) which cannot be changed by a diffeomorphism
(using Mostow rigidity) and secondly one gets a codimension-one foliation
of $Y$ with Godbillon-Vey number proportional to the square of the
scale, i.e. $GV(Y)\sim r^{2}$. Now we will extend these properties
to every member of the radial family. The strategy of the proof goes
like this: we use the foliation of one member (using theorem \ref{thm:foliation-single-small-exotic-R4})
to induce a foliation of the design of the Casson handle (see previous
subsection) for the radial family $\mathbb{R}_{t}^{4}$. Then we will
obtain a foliation of every member $R_{t}^{4}$ with $t\in CS\subset[0,1]$.
By using lemma \ref{lem:foliation-gap-1}, we will obtain also a foliation
of the gaps and therefore for every member of the radial family. The
codimension-one foliations of $Y_{r}$ (surrounding the member $R_{t}^{4}$
with $r=\frac{1}{1-t}$) has a non-trivial Godbillon number proportional
to $r^{2}=\frac{1}{(1-t)^{2}}$. Importantly, the whole constructions
works only for a fixed radial family $\mathbb{R}_{t}^{4}$, i.e. a
renumbering of the parameter $t$ does not produce new exotic $\mathbb{R}^{4}$
in the radial family or the radius $r$ as well the parameter $t$
is not a diffeomorphism invariant. In physics, we need only a fixed
radial family and every statement will be understood relative to this
family. \begin{theorem} \label{thm:codim-1-foli-radial-fam} A fixed
radial family $\mathbb{R}_{t}^{4}$ of small exotic $\mathbb{R}^{4}$'s
, i.e. of members $R_{t}^{4}$ with radius $r$ and $t=1-\frac{1}{r}$,
is constructed from the non-product h-cobordism $W$ between $M$
and $M_{0}$ with Akbulut corks $A\subset M$ and $A\subset M_{0}$,
respectively. Let $Y_{r}$ be a 3-manifold surrounding the member
$R_{t}^{4}$ with $t=1-\frac{1}{r}$ (in the notation above). Every
member $R_{t}$ of the radial family determines a codimension-one
foliations of $Y_{r}$ with Godbillon-Vey number $r^{2}$. Furthermore
given two members $R_{t}$ and $R_{s}$ with $s\not=t$, then the
two members are non-isotopic and the two corresponding codimension-one
foliations of $Y_{r}$ and $Y_{u}$, respectively, are non-cobordant
to each other.\end{theorem} \emph{Proof:} In \cite{DeMichFreedman1992},
this radial family is described by a sequence of embeddings $R_{0}^{4}\subset R_{t}^{4}\subset R_{1}^{4}$
(see Fig. \ref{fig:Radial-family-exoticR4}). Furthermore, there is
a compact subset $K\subset R_{0}^{4}$ which cannot be surrounded
by a 3-sphere. Therefore every member of the radial family is a small
exotic $\mathbb{R}^{4}$. The member $R_{1}^{4}$ was explicitly constructed
in \cite{BizGom:96} using the simplest tree $\mathcal{T}_{+}$ (see
above). Every tree $\mathcal{T}$ having $\mathcal{T}_{+}$ as subtree
defines a Casson handle $CH_{\mathcal{T}}$ with $CH_{\mathcal{T}}\subset CH_{\mathcal{T}_{+}}$.
The class of all trees with this property defines a family of $R_{t}^{4}$
indexed by the Cantor set $CS\subset[0,1]$. Every member of this
family admits a foliation according to theorem \ref{thm:foliation-single-small-exotic-R4}
and the size is controlled by the lemmas \ref{lem:size-control-Y}
and \ref{lem:foliation-Y-for-member} so that every member $R_{t}^{4}$
with $t\in CS$ induces a foliation on $Y_{r}$ with $t=1-\frac{1}{r}$
and Godbillon-Vey number $GV(Y_{r})=r^{2}$. Now we have to discuss
the members $R_{t}^{4}$ with $t\not\in CS$. Using lemma \ref{lem:foliation-gap-1},
the foliation of the design (the $CS$ indexed subset) induces a foliation
of the gaps which will be used now to extend the foliation to every
member of the radial family. The gaps in case of the Casson handle
are not manifolds and look like $S^{1}\times D^{3}/Wh$. In case of
the capped grope one has ``good'' gaps of the form $S^{1}\times D^{3}$.
That is the reason why we switch to these objects now. Now we decompose
the gap by $gap=S^{1}\times D^{3}=S^{1}\times D^{2}\times I$ with
the unit interval $I=[0,1]=D^{1}$. The boundary is a decomposition
$\partial(gap)=(S^{1}\times S^{1}\times I)\cup(S^{1}\times D^{2}\times\left\{ 0,1\right\} )$
of the caps (north and south) and the equator region. We remark the
importance of the boundary of the gap (see the proof of lemma \ref{lem:foliation-gap-1}).

The radius coordinate $\rho$ defined above is identified with the
unit interval of the $gap$ (see the proof of Theorem 3.2 in \cite{DeMichFreedman1992}).
In the notation of \cite{DeMichFreedman1992}, we think of each gap
as $gap=S^{1}\times N\times I$ where $N=D^{2}$ is the neighborhood
around the north pole of the 2-sphere. Using the reembedding theorems
every GCH embeds in the open 2-handle and induces a foliation. As
described above the simplest tree $\mathcal{T}_{+}$ belongs to the
binary sequence $.000\ldots$ and is represented by $t=1$ and the
radius $r=1/(1-t)=+\infty$. The foliation of the design is perpendicular
to $S^{1}\times N$, i.e. $S^{1}\times\{latitude\}$ are the leaves.
The intersection of the leaves with $S^{1}\times N$ produces a foliation
of the disk $N$ as proved in theorem \ref{lem:foliation-gap-1}.
This disk is given up to conformal automorphism by fixing the sphere
$S^{2}\supset N$, i.e. the disk is invariant w.r.t. the group $PSL(2,\mathbb{R})$.
The boundary of $N$ is given by geodesic curves. The $PSL(2,\mathbb{R})-$invariance
induces a mapping of the disk $N$ into the hyperbolic space $\mathbb{H}^{2}$,
where every $PSL(2,\mathbb{R})$ transformation is an isometry now.
Then the foliated $N$ is mapped to a foliated polygon $P$ in $\mathbb{H}^{2}$,
where the foliation is $PSL(2,\mathbb{R})-$invariant. From this point
of view we interpret $S^{1}\times N$ as the unit tangent bundle $UP$
of the polygon $P$. Then the volume of the polygon $P$ is the volume
of the disk $N$, i.e. $vol(P)=vol(N)$ and we choose the number of
vertices of $P$ in a suitable manner by defining the geodesic arcs
forming the boundary of $N$. The disk $N$ is also part of the boundary
$\partial(gap)=S^{1}\times S^{2}$ of the gap using its foliation,
see lemma \ref{lem:foliation-gap-1}. Then the unit interval in the
gap is directly related to the radius $r$ of the 2-sphere $S^{2}\supset N$
and this radius determines the volume of the disk $N$ (as part of
the upper hemisphere of $S^{2}$. But then by using $PSL(2,\mathbb{R})-$invariance,
we obtain the relation $vol(P)=r^{2}$. Therefore, every member $R_{t}^{4}$
of the radial family (i.e. $t\in[0,1]$) is surrounded by $Y_{r}$
with $t=1-\frac{1}{r}$ and admits a codimension-one foliation with
Godbillon-Vey number $GV(Y_{r})=r^{2}$.

Let $R_{t}^{4}$ and $R_{s}^{4}$ be two members of the radial family
with $s\not=t$ surrounded by 3-manifolds $Y_{r}$ and $Y_{s}$ with
radii $r=\frac{1}{1-t}$ and $u=\frac{1}{1-s}$, respectively. By
Theorem 3.2 in \cite{DeMichFreedman1992}, there are two possible
cases for $s,t\in CS$: $R_{s}^{4}$ and $R_{t}^{4}$ are non-diffeomorphic
(a subset of $CS$ of continuous cardinality) or both spaces are diffeomorphic
but non-isotopic (different Casson handles are non-isotopic). But
by the construction of radial family, see \cite{DeMichFreedman1992},
one knows that every member is constructed by some Casson handle (some
of them index by $CS$ and some of them lying in the gaps). So that
we obtain the corresponding Casson handles $CH_{s}$ and $CH_{t}$.
But $s\not=t$ means $CH_{s}$ is different from $CH_{t}$, so at
least non-isotopic (see subsection \ref{sub:Isotopy-of-surfaces}
for a discussion). Therefore the foliations of $Y_{u}$ and $Y_{r}$
with $GV(Y_{u})=u^{2}$ and $GV(Y_{r})=r^{2}$ corresponding to at
least non-isotopic $R_{s}^{4}$ and $R_{t}^{4}$. Or, the Godbillon-Vey
numbers label the isotopy classes of the members (non-diffeomorphic
members are also non-isotopic). The corresponding foliations are non-cobordant
to each other. $\square$

Here are some remarks. At first the theorem \ref{thm:foliation-single-small-exotic-R4}
can be extended to every $Y_{n}$ separating the compact subset from
infinity in the member $R_{t}^{4}$. Secondly by using theorem \ref{thm:foliation-3MF}
we can trace back the foliation on $Y_{r}$ to a foliation on the
3-sphere $S^{3}$. In both 3-manifold, there is a submanifold which
looks like $V^{2}\times S^{1}$ (in the notation above) and we will
foliate $V^{2}\times S^{1}$ with non-zero Godbillon-Vey number. We
will say \emph{that $S^{3}$ lies at the boundary $\partial\overline{R_{t}^{4}}=Y_{r}$}
\emph{of the completion of the small exotic $R_{t}^{4}$}. But we
remark there is actually no 3-sphere, otherwise we obtain a contradiction
to the exoticness. Then by theorem \ref{thm:codim-1-foli-radial-fam}:
\begin{corollary} Any class in $H^{3}(S^{3},\mathbb{R})$ induces
a small exotic $\mathbb{R}^{4}$ where $S^{3}$ lies at the boundary
of the completion of the small exotic $\mathbb{R}^{4}$ (using an
embedding in the 4-sphere). \end{corollary} We will understand the
role of the 3-sphere after the next section where we discuss the small
exotic $\mathbb{R}^{4}$ as a deformation of the standard $\mathbb{R}^{4}$.

\section{Small exotic $\mathbb{R}^{4}$'s as deformed standard $\mathbb{R}^{4}$\label{sec:deformation-standard-R4}}

\begin{figure}
\centerline{\psfig{file=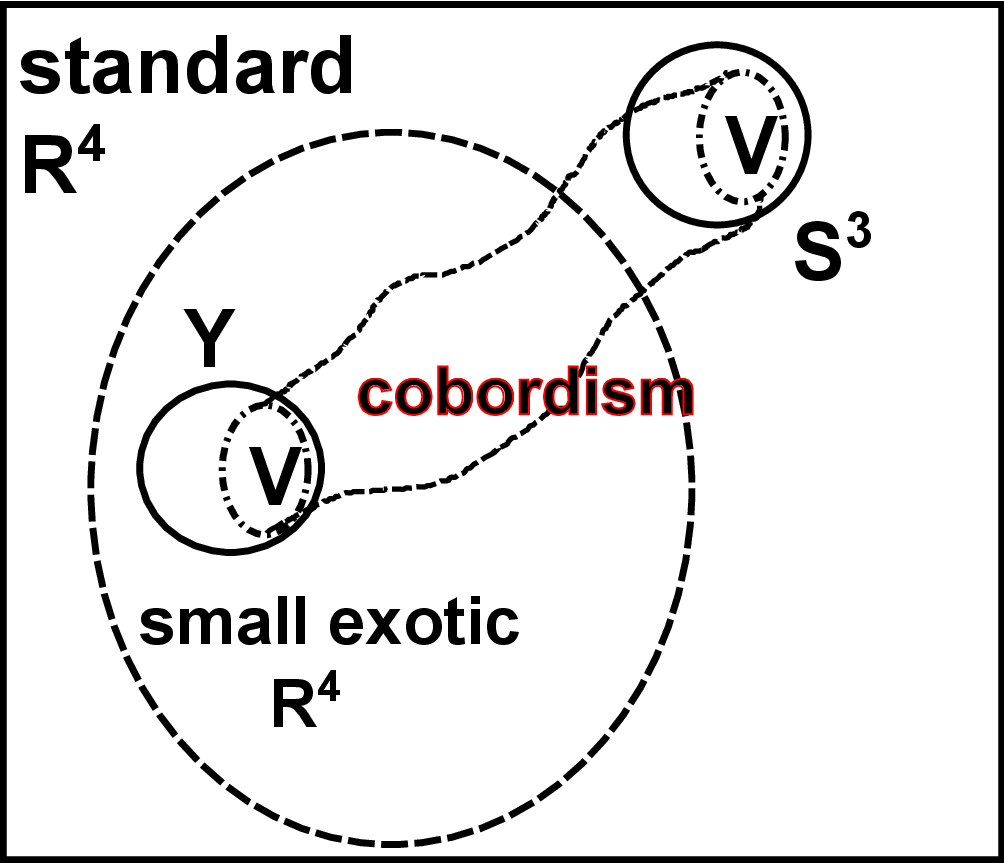,width=6cm}}
\caption{Schematic picture of the cobordism between the submanifolds $V\subset Y_{n}\subset R^{4}$
and $V\subset S^{3}\subset\mathbf{R}^{4}$ (the 3-sphere can also
surround $R^{4}$)\label{fig:Schematic-picture-of-cobordism-Y-S3}}
\end{figure}

Let $\mathbf{R}^{4}$ be the standard $\mathbb{R}^{4}$ and $R^{4}$
a small exotic $\mathbb{R}^{4}$. For the following construction (see
\cite{DeMichFreedman1992}), it is enough to discuss the special case
\cite{BizGom:96}. The main idea is the construction of appendix A.
Let $\mathcal{T}$ be a tree then one can construct a variant of the
Whitehead continuum $WhC_{\mathcal{T}}$ and the open 4-manifold 
\[
U_{\mathcal{T}}=D^{2}\times D^{2}\setminus\mbox{cone}(WhC_{\mathcal{T}})
\]
so that $U_{\mathcal{T}}$ is diffeomorphic to the Casson handle $CH_{\mathcal{T}}$
(relative to the attaching region $\partial D^{2}\times D^{2}$).
In Fig. \ref{fig:link-picture-for-K}, we have a description of the
compact subset $K\subset R^{4}$. If we use the standard 2-handle
$D^{2}\times D^{2}$ then we obtain the 4-disk $D^{4}$ (the 1- and
2-handle in Fig. \ref{fig:link-picture-for-K} cancel each other).
Therefore we obtain 
\[
\left(K\cup(D^{2}\times D^{2})\right)\setminus\mbox{cone}(WhC_{\mathcal{T}})=D^{4}\setminus\mbox{cone}(WhC_{\mathcal{T}})
\]
and the interior of this 4-manifold is the small exotic $\mathbb{R}^{4}$
called $R^{4}$. Thus we have the rule:\\
 \textbf{Rule}: Remove the boundary $\partial D^{4}=S^{3}$ and the
space $\mbox{cone}(WhC_{\mathcal{T}})$, then you will get the $R^{4}$.
Alternatively, remove the subset $\mbox{cone}(WhC_{\mathcal{T}})$
from $\mathbf{R}^{4}$ and you will obtain $R^{4}$. We call this
procedure:\emph{ the deformation of the standard $\mathbf{R^{4}}$
to get a small exotic $\mathbb{R}^{4}$.}\\
 But at the same time, $R^{4}$ embeds into $\mathbf{R}^{4}$ (see
Fig. \ref{fig:Radial-family-exoticR4}) and the compact subset $K$
is separated from infinity by the 3-manifold $Y_{n}$. Furthermore,
$R^{4}$ is surrounded by the 3-manifold $Y_{\infty}\subset\mathbf{R}^{4}$.
We know by Theorem \ref{thm:foliation-single-small-exotic-R4} that
$Y_{\infty}$ but also $Y_{n}$ admits a codimension-one foliation
with non-zero Godbillon-Vey number. Both 3-manifolds $Y_{\infty}$
and $Y_{n}$ are submanifolds of $\mathbf{R}^{4}$ at the same time.
But then, a neighborhood $U(Y_{n})\subset\mathbf{R}^{4}$ is surrounded
by a 3-sphere $S^{3}=\partial(\overline{U(Y_{n})})$ using the properties
of the standard $\mathbf{R}^{4}$. By using the decomposition (\ref{eq:decomposition-3MF-Y})
of $Y_{n}$ and (\ref{eq:decomposition-S3}) of $S^{3}$, we obtain
a common submanifold $V^{2}\times S^{1}$ of $Y_{n}$ and $S^{3}$.
By construction of the codimension-one foliation using Theorem \ref{thm:foliation-3MF},
there is a foliated cobordism between $(V^{2}\times S^{1})_{Y}\subset Y_{n}$
and $(V^{2}\times S^{1})_{S^{3}}\subset S^{3}$ so that the foliation
of $Y_{n}$ goes over to the 3-sphere $S^{3}$. In particular, both
Godbillon-Vey numbers are equal. Therefore we showed: \begin{theorem}
\label{thm:foliation-S3-from-Y} Let $\mathbb{I}:R^{4}\hookrightarrow\mathbf{R}^{4}$
be an embedding of the small exotic $\mathbb{R}^{4}$ into the standard
$\mathbb{R}^{4}$. As constructed above, $Y_{n}\subset R^{4}$ separates
$K\subset R^{4}$ from infinity. Using the embedding $\mathbb{I}$,
there is a neighborhood $U(Y_{n})\subset\mathbf{R}^{4}$ which is
surrounded by a 3-sphere $S^{3}=\partial(\overline{U(Y_{n})})\subset\mathbf{R}^{4}$.
Then the codimension-one foliation of $Y_{n}\subset R^{4}$ (from
$R^{4}$) induces (by a foliated cobordism) a codimension-one foliation
of $S^{3}$ with equal Godbillon-Vey numbers. \end{theorem} Properties
of $R^{4}$ encoded into the foliation of $Y_{n}$ go over to the
foliation of the 3-sphere $S^{3}$ in the standard $\mathbf{R}^{4}$
(see Fig. \ref{fig:Schematic-picture-of-cobordism-Y-S3}). 

Now we are interested in an expression of this deformation at the
level of functions, geometries etc. using the rule above, i.e. 
\[
R^{4}=\mathbf{R}^{4}\setminus\mbox{cone}(WhC_{\mathcal{T}})
\]
and for the 3-sphere at infinity we have 
\[
C(WhC_{\mathcal{T}})=S_{\infty}^{3}\setminus WhC_{\mathcal{T}}
\]
the complement of the Whitehead continuum, a wild embedded 3-manifold.
In \cite{AsselmeyerKrol2013}, we discussed wild embedded submanifolds
as a geometric expression of a quantum state. Here we will use it
to obtain the deformation of the algebra of all complex functions
to a noncommutative $C^{*}-$algebra, the space of holonomies along
closed curves (as an expression of geometry and its algebra of observables)
to the skein module and the deformation of the Levi-Civita connection
to a Hitchin structure. We will give only a brief account to these
problems which will be expanded in our forthcoming work. It is a motivation
for the following sections.

\textbf{1.} \textbf{Function algebra}: Let us consider the algebra
$C_{0}(S^{3})$ of complex functions (with compact support) over the
3-sphere. The complement of a (tame) contractable, connected subset
of the 3-sphere is always simply-connected. But the complement $C(WhC_{\mathcal{T}})=S^{3}\setminus WhC_{\mathcal{T}}$
of the Whitehead continuum is non-simply-connected, i.e. the fundamental
group $\pi_{1}(C(WhC_{\mathcal{T}}))$ is non-zero. Then the automorphism
group of $C(WhC_{\mathcal{T}})$ consists of two parts: the diffeomorphism
group $Diff(S^{3})$ and $Aut(\pi_{1}(C(WhC_{\mathcal{T}})))$ put
together by a semi-direct product. This situation is well-known in
noncommutative geometry \cite{Connes94}. We will understand the deformation
of the function algebra $C_{0}(S^{3})$ as the following construction
of a $C^{*}-$algebra (see\cite{AsselmeyerKrol2013}). At first we
note that the construction of wild embeddings is usually given by
an infinite construction%
\footnote{This infinite construction is necessary to obtain an infinite polyhedron,
the defining property of a wild embedding.%
}. From an abstract point of view, we have a decomposition of $\mathcal{G}=C(WhC_{\mathcal{T}})=S^{3}\setminus WhC_{\mathcal{T}}$
by an infinite union 
\[
\mathcal{G}=\bigcup_{i=0}^{\infty}C_{i}
\]
of 'level sets' $C_{i}$. Then every element $\gamma\in\pi=\pi_{1}(C(WhC_{\mathcal{T}}))$
lies (up to homotopy) in a finite union of levels. The basic elements
of the $C^{*}-$algebra $C^{*}(\mathcal{G},\pi$) are smooth half-densities
with compact supports on $\mathcal{G}$, $f\in C_{c}^{\infty}(\mathcal{G},\Omega^{1/2})$,
where $\Omega_{\gamma}^{1/2}$ for $\gamma\in\pi$ is the one-dimensional
complex vector space of maps from the exterior power $\Lambda^{k}L$
($\dim L=k$), of the union of levels $L$ representing $\gamma$,
to $\mathbb{C}$ such that 
\[
\rho(\lambda\nu)=|\lambda|^{1/2}\rho(\nu)\qquad\forall\nu\in\Lambda^{2}L,\lambda\in\mathbb{R}\:.
\]
For $f,g\in C_{c}^{\infty}(\mathcal{G},\Omega^{1/2})$, the convolution
product $f*g$ is given by the equality 
\[
(f*g)(\gamma)=\intop_{\gamma_{1}\circ\gamma_{2}=\gamma}f(\gamma_{1})g(\gamma_{2})
\]
with the group operation $\gamma_{1}\circ\gamma_{2}$ in $\pi$. Then
we define via $f^{*}(\gamma)=\overline{f(\gamma^{-1})}$ a $*$operation
making $C_{c}^{\infty}(\mathcal{G},\Omega^{1/2})$ into a $*$algebra.
Each level set $C_{i}$ consists of simple pieces (in case of the
Whitehead continuum the pieces are solid tori) denoted by $T$. For
these pieces, one has a natural representation of $C_{c}^{\infty}(\mathcal{G},\Omega^{1/2})$
on the $L^{2}$ space over $T$. Then one defines the representation
\[
(\pi_{x}(f)\xi)(\gamma)=\intop_{\gamma_{1}\circ\gamma_{2}=\gamma}f(\gamma_{1})\xi(\gamma_{2})\qquad\forall\xi\in L^{2}(T),\forall x\in\gamma.
\]
The completion of $C_{c}^{\infty}(\mathcal{G},\Omega^{1/2})$ with
respect to the norm 
\[
||f||=\sup_{x\in\mathcal{G}}||\pi_{x}(f)||
\]
makes it into a $C^{*}$algebra $C_{c}^{\infty}(\mathcal{G},\pi$).
We will call this $C^{*}$algebra, the \emph{deformation of the function
algebra}.

\textbf{2. Geometry via holonomy}: A geometry can be defined in many
different ways. One way as pioneered by Cartan is the setting of an
isometry group $G$ so that the space $N$ is the factor space $G/H$
for a subgroup $H$ at least locally. In case of 3-manifolds, one
can split the 3-manifold into pieces so that every pieces is a factor
space of this kind (Thurstons Geometrization Theorem). To determine
the subgroup $H$, one needs the holonomy along a closed curve or
the representation $\pi_{1}(N)\to G$ of the fundamental group. By
the Ambrose-Singer theorem, the holonomy determines the curvature
uniquely. Furthermore, the holonomy is an important variable of a
geometry. Then we can form the space 
\[
X(N,G)=\frac{Hom(\pi_{1}(N),G)}{G}
\]
the space of all representation $\pi_{1}(N)\to G$ (homomorphisms)
up to conjugacy. The functions over this space are the set of observables
of our geometric theory. For the 3-sphere, $X(S^{3},G)$ is trivial
and we recover the geometry with positive constant curvature for $S^{3}$.
In case of a surface $S$, the space $X(S,G)$ admits a Poisson structure
(see \cite{Skovborg2006}) or the (complex) functions over $X(S,G)$
form a Poisson algebra. Now following \cite{AsselmeyerKrol2013},
the generators of the fundamental group $\pi_{1}(C(WhC_{\mathcal{T}}))$
are given by non-contractable curves around the solid tori forming
the Whitehead continuum. The boundary of a solid torus is the usual
torus, i.e. a surface of genus one. Therefore like in \cite{AsselmeyerKrol2013}
we can use the Poisson structure of $X(S,G)$ where $S$ are tori.
Then the wild embedding is given by a deformation quantization, i.e.
the algebra of observables (functions over $X(S,G)$) is changed to
the skein module over $S\times[0,1]$. We end up with the space of
singular knots over $S\times[0,1]$ (as configuration space of the
quantized theory) together with a line bundle so that a section in
this bundle is the wave function. This situation was described in
\cite{Brylinski1993} and was a main motivation for the introduction
of a gerbe. In this manner, the geometry of $S^{3}$ (as boundary
of the $\mathbf{R}^{4}$) is deformed to a complicated geometry for
$C(WhC_{\mathcal{T}})$ (as boundary of the exotic $R^{4}$) given
by a gerbe over $S^{3}$.

3. \textbf{Vector fields and 1-forms}: Now we will discuss the deformation
of vector fields and 1-forms by the change $\mathbf{R}^{4}$ to $R^{4}$.
We will see that it is related to the deformation of a Hitchin structure
by a 3-form. As explained in the proof of Theorem \ref{thm:codim-1-foli-radial-fam}
(using subsection \ref{sub:Isotopy-of-surfaces}), the difference
between $\mathbf{R}^{4}$ and $R^{4}$ is at least given by different
isotopy classes of embeddings of $Y_{n}$. Here we will use a method
to detect isotopy classes by the Chern-Simons invariant $CS(A)$ of
the tangent bundle $TY_{n}$ (with connection $A$). The functional
\[
CSF(A)=\intop_{Y_{n}}tr(A\wedge dA+\frac{2}{3}A^{3})
\]
of $CS(A)=CSF(A)\bmod\mathbb{Z}$ is only invariant for small diffeomorphisms
(coordinate transformations) and changes by an integer (with an additional
factor) for large diffeomorphisms. The Chern-Simons invariant can
be related to another invariant, the eta invariant $\eta$ used in
the index theory for manifolds \cite{AtiPatSin:73}. One obtains 
\[
CS(A_{\rho})=3\eta_{\rho}\bmod\mathbb{Z}
\]
for the Levi-Civita connection $A_{\rho}$ of a metric $\rho$ (where
the eta invariant belongs to the signature complex). Here the Chern-Simons
invariant is at a special branch (acquiring its minimal value, see
\cite{FinSte:90}). The difference 
\[
CS(A)-3\eta_{\rho}
\]
defines the shift for large diffeomorphisms, i.e. for non-trivial
isotopy classes. Using \cite{Atiyah1990b}, this difference is the
2-framing $\sigma(\alpha)$ of $Y_{n}$ 
\[
\sigma(\alpha)=CS(A)-3\eta_{\rho}
\]
with respect to a trivialization $\alpha$ (up to homotopy) of the
vector bundle 
\[
TY_{n}\oplus TY_{n}
\]
canonically equivalent to 
\[
TY_{n}\oplus T^{*}Y_{n}
\]
Geometric structures on this bundle are known as Hitchin structures
and the 3-form $tr(A\wedge dA+\frac{2}{3}A^{3})$ is the corresponding
deformation. In the spirit of Theorem \ref{thm:foliation-S3-from-Y},
one can start with a Hitchin structure on $TS^{3}\oplus T^{*}S^{3}$
and deform it by a 3-form to get a non-trivial isotopy class belonging
to non-trivial embedded $S^{3}$ which is again related to a non-trivial
embedded $Y_{n}$ (with non-trivial isotopy class). This argumentation
motivates the introduction of deformed Hitchin structures (as the
deformation of vector fields and 1-forms for exotic $R^{4}$).

\section{Abelian gerbes on $S^{3}$\label{sub:U(1)--gerbes-on}}

As we saw in the previous section, the geometry of the exotic $R^{4}$
can be understood by line bundles over the loop space or by gerbes.
The gerbes are related to foliations with integer Godbillon-Vey numbers.

\subsection{Integer Godbillon-Vey invariants and flat bundles\label{sub:Integer-Godbillon-Vey-invariants}}

All results in this section are well-known but we will present them
for completeness. Clearly the integer classes $H^{3}(S^{3},\mathbb{Z})\subset H^{3}(S^{3},\mathbb{R})$
are a subset of the full set (induced by embedding $\mathbb{Z}\to\mathbb{R}$)
and one can use the construction above to get the foliation. Using
the work of Goldman and Brooks \cite{BrooksGoldman1984}, one can
also construct a foliation admitting an integer Godbillon-Vey invariant.
The corresponding foliation is again induced by the unit tangent bundle
$U\mathbb{H}^{2}$ or by the action of the M{\"o}bius group $PSL(2,\mathbb{R})=SL(2,\mathbb{R})/\mathbb{Z}_{2}$
(Remark: $PSL(2,\mathbb{R})$ acts transitively on $U\mathbb{H}^{2}$
and so we can identify both spaces). The unit tangent bundle $U\mathbb{H}^{2}=PSL(2,\mathbb{R})$
is a circle bundle over $\mathbb{H}^{2}$ and we can construct the
universal cover, a real line bundle over $\mathbb{H}^{2}$, denoted
by $\widetilde{SL}(2,\mathbb{R})$. In subsection \ref{sub:Non-cobordant-foliations-S3}
we described the construction of a codimension-one foliation $\mathcal{F}$.
In an intermediate step one has the manifold $M=V^{2}\times S^{1}=(S^{2}\setminus\left\{ \mbox{\mbox{k} punctures}\right\} )\times S^{1}$
(with a foliation $\mathcal{F}$). This foliation $\mathcal{F}$ is
defined by a one-form $\omega$ ($PSL(2,\mathbb{R})$ invariant) together
with two other 1-forms $\theta,\eta$ with 
\begin{equation}
d\omega=\theta\wedge\omega,\quad d\theta=\omega\wedge\eta,\quad d\eta=\eta\wedge\theta\label{eq:sl2-relations}
\end{equation}
and Godbillon-Vey invariant $GV(\mathcal{F})=\theta\wedge d\theta=\omega\wedge\eta\wedge\theta$.
Now we show that the Godbillon-Vey invariant of this foliation $\mathcal{F}$
is an integer 3-form: \begin{lemma} \label{lem:integer-GV}Let $M$
be a 3-manifold with non-trivial fundamental group $\pi_{1}(M)$ and
a foliation $\mathcal{F}$ defined by the 1-form $\omega$ together
with two 1-forms $\theta,\eta$ fulfilling the relations (\ref{eq:sl2-relations}).
If $M$ can be written as a flat $PSL(2,\mathbb{R})-$bundle over
a manifold $N$ with fiber $S^{1}$ and $\pi_{1}(N)\not=0$. Then
the pairing of the Godbillon-Vey invariant $\Gamma_{\mathcal{F}}$
with the fundamental class $[M]\in H_{3}(M)$ is given by 
\begin{equation}
GV(M,\mathcal{F})=\langle\Gamma_{\mathcal{F}},[M]\rangle=\intop_{M}\Gamma_{\mathcal{F}}=\pm(4\pi)^{2}\cdot\chi(N)\label{eq:integer-GV}
\end{equation}
with the Euler characteristic $\chi(N)$ of $N$. Up to a normalization
constant and the orientation one obtains an integer value.\end{lemma}
\emph{Proof:} According to the Thurston's construction, we choose
$M=(S^{2}\setminus\left\{ \mbox{\mbox{k} punctures}\right\} )\times S^{1}$
and $N=S^{2}\setminus\left\{ \mbox{\mbox{k} punctures}\right\} $.
The group $PSL(2,\mathbb{R})$ is the isometry group of the hyperbolic
plane $\mathbb{H}^{2}$ leaving the foliation of $M$ invariant. Then
(see \cite{BrooksGoldman1984}) the holonomy of the foliation is given
by a homomorphism $\pi_{1}(M)\to\widetilde{SL}(2,\mathbb{R})$ defining
a flat bundle over $N$. Using Proposition 2 of \cite{BrooksGoldman1984},
the integration of the Godbillon-Vey invariant over the fiber $S^{1}$
gives 
\[
\intop_{S^{1}}GV(\mathcal{F})=(4\pi)^{2}\cdot e(M)
\]
with the Euler class $e(M)$ of the flat bundle $M$. An integration
over $N$ gives by definition the Euler characteristics $\chi(N)$.
Thus we obtain the desired result: 
\[
GV(M,\mathcal{F})=\langle\Gamma_{\mathcal{F}},[M]\rangle=\intop_{M}\Gamma_{\mathcal{F}}=(4\pi)^{2}\cdot\chi(N)
\]
the integer invariant. A change of the orientation changes the sign
of the integral. Then we obtain the negative integers. $\square$

Using this lemma we are able to obtain the foliation of the $S^{3}$
with integer Godbillon-Vey invariant. Putting the above results together,
we obtain: \begin{theorem} \label{thm:integer-GV-flat-bundles}Every
$PSL(2,\mathbb{R})$ flat bundle over $M=(S^{2}\setminus\left\{ \mbox{\mbox{k} punctures}\right\} )\times S^{1}$
defines a codimension-one foliation of $M$ by the horizontal distribution
of the flat connection so that its (normalized) Godbillon-Vey invariant
is an integer given by 
\begin{equation}
\frac{1}{(4\pi)^{2}}\langle\Gamma_{\mathcal{F}},[M]\rangle=\pm\chi(N)=\pm(2-k)\quad.\label{eq:integer-GV-S3}
\end{equation}
This foliation can be extended to the whole 3-sphere $S^{3}$ defining
an integer class in $H^{3}(S^{3},\mathbb{Z})$.\end{theorem} It is
an important consequence of the work \cite{BrooksGoldman1984} that
the foliation $\mathcal{F}$ (and its induced counterpart for the
3-sphere $S^{3}$) is rigid, i.e. a disturbance (or continuous variation)
does not change the Godbillon-Vey invariant.

\subsection{Gerbes and small exotic $\mathbb{R}^{4}$'s}

In this subsection we will focus on the relation of the foliation
with integer Godbillon-Vey number to gerbes. Abelian gerbes and gerbe
bundles on manifolds are geometrical objects interpreting third integral
cohomologies on manifolds. From our point of view, there are two interpretations
for classes in $H^{3}(M,\mathbb{Z})$: as (integer) Godbillon-Vey
classes and (as we will see) as abelian gerbes. There exists a vast
literature in mathematics and physics devoted to gerbes and bundle
gerbes. Gerbes were first considered by Giraud \cite{Giraud1971}.
The classical reference is the Brylinski book \cite{Brylinski1993}.
Here we are interested in abelian gerbes. Up to uniqueness questions
of the choices we can, following Hitchin, make use of the simplified
working definition of a gerbe \cite{Hitchin1999}. Hence we do not
refer to categorical constructions (sheaves of categories, see e.g.
\cite{Murray1996}) which, from the other hand, are essential for
the correct recognition of gerbes. In that way we can state easily
the relation of gerbes to 3-rd integral cohomologies on $S^{3}$.

Abelian, or $S^{1}$, gerbes are best understood in terms of cocycles
and corresponding transition objects. As warm-up, a $S^{1}$ principal
bundle over a manifold $M$ is specified by a cocycle $g_{\alpha\beta}:U_{\alpha}\cap U_{\beta}\to S^{1}$
which is the ${\rm \breve{C}}$ech cocycle in $\breve{C}^{1}(M,C^{\infty}(S^{1}))$
where $U_{\alpha,\beta}$ are elements of a good cover of a manifold
$M$. In contrast, to define a $S^{1}$ gerbe, we need to compare
data on each triple intersections of elements w.r.t. a good cover.
Hence, let $g_{\alpha\beta\gamma}:U_{\alpha}\cap U_{\beta}\cap U_{\gamma}\to S^{1}$
be a cocycle in $\breve{C}^{2}(M,C^{\infty}(S^{1}))$ with

\begin{equation}
g_{\alpha\beta\gamma}=g_{\beta\alpha\gamma}^{-1}=g_{\beta\gamma\alpha}^{-1}=g_{\alpha\gamma\beta}^{-1}\label{eq:gerbeg}
\end{equation}
satisfying the cocycle condition

\begin{equation}
\delta g=g_{\beta\gamma\delta}g_{\alpha\gamma\delta}^{-1}g_{\alpha\beta\delta}g_{\alpha\beta\gamma}^{-1}=1\label{eq:gerbecocycle}
\end{equation}
on each fourth intersection $U_{\alpha}\cap U_{\beta}\cap U_{\gamma}\cap U_{\delta}$.
Then this cocyle defines the \emph{$S^{1}-$gerbe}.

Then the classification of $S^{1}$gerbes is similar like $S^{1}$
principal bundles. The cocycles in $\breve{C}^{2}(M,C^{\infty}(S^{1}))$
give rise (up to the coboundaries) to the cohomology group $H^{2}(M,C^{\infty}(S^{1}))$
which classifies gerbes as defined above. However, we have the canonical
exact sequence of sheaves on $M$

\begin{equation}
0\longrightarrow\mathbb{Z}\longrightarrow C^{\infty}(\mathbb{R})\longrightarrow C^{\infty}(S^{1})\longrightarrow1
\end{equation}
where the third morphism is given by $e^{2\pi ix}$. However, $C^{\infty}(\mathbb{R})$
is fine, hence

\begin{equation}
H^{2}(M,C^{\infty}(S^{1}))=H^{3}(M,\mathbb{Z})
\end{equation}
We see that gerbes are classified by elements in the third integral
cohomology on $M$ similarly as line bundles are classified topologically
by Chern classes as elements in the second cohomology. The elements
of $H^{3}(M,\mathbb{Z})$ are called the Dixmier-Douady classes of
the ``gerbe local data''. It is worth noticing that gerbes are neither
manifolds nor bundles. They can be considered as a generalization
of both: sheaves (bundle gerbes) and vector bundles (cocycle description).

\emph{A trivialization} of a $S^{1}$- gerbe is given by functions

\begin{equation}
f_{\alpha\beta}=f_{\beta\alpha}^{-1}:U_{\alpha}\to U_{\beta}\label{eq:triv.gerb}
\end{equation}
such that

\begin{equation}
g_{\alpha\beta\gamma}=f_{\alpha\beta}f_{\beta\gamma}f_{\gamma\alpha}
\end{equation}
which is a representation of a cocycle by functions. Thus the difference
of two trivializations is given by $h_{\alpha\beta}=f_{\alpha\beta}/f'_{\alpha\beta}$
which means $h_{\alpha\beta}=h_{\beta\alpha}^{-1}$ and

\begin{equation}
h_{\alpha\beta}h_{\beta\gamma}h_{\gamma\alpha}=1
\end{equation}
and this is exactly a cocycle for some line bundle on $M$. We say
that the (generalized) transition functions of an abelian gerbe are
line bundles. One can iterate this construction and define higher
,,gerbes'' with the generalized transition functions given by lower
rank gerbes.

\emph{A connection} on a gerbe (\emph{local data}) is specified by
1-forms $A_{\alpha\beta}$ and 2-forms $B_{\alpha}$ satisfying the
following two conditions

\begin{equation}
iA_{\alpha\beta}+iA_{\beta\gamma}+iA_{\gamma\alpha}=g_{\alpha\beta\gamma}^{-1}dg_{\alpha\beta\gamma}\label{eq:gerbeConnection1}
\end{equation}

\begin{equation}
B_{\beta}-B_{\alpha}=dA_{\alpha\beta}\label{eq:gerbeConnection2}
\end{equation}
This implies that there exists a globally defined 3-form $H$, with
integral 3-rd deRham cohomologies corresponding to $[H/2\pi]$ and
defined by \emph{local} 2-forms $B_{\alpha}$

\begin{equation}
H|_{U_{\alpha}}=dB_{\alpha}
\end{equation}
This 3-form $H$ is the \emph{curvature} of a gerbe with connection
as above. The local data defining a gerbe exists whenever $[H/2\pi]$
is integral.

The trivialization $f_{\alpha\beta}$ is the element of $\breve{C}^{1}(M,C^{\infty}(S^{1}))$
so that 

\begin{equation}
[g_{\alpha\beta\gamma}]=0
\end{equation}
for trivial gerbe. The connection on a gerbe is \emph{flat} when $H|_{U_{\alpha}}=dB_{\alpha}=0$
for a suitably chosen cover $\{U_{\alpha}\}$. In that case we have
$B_{\alpha}=da_{\alpha}$ for the suitably chosen cover.

In case of $M=S^{3}$ the integral third deRham cohomology group (as
image of the map $H^{3}(S^{3},\mathbb{R})\to H^{3}(S^{3},\mathbb{Z})$)
classifies $S^{1}$-gerbes on $S^{3}$. These integral classes are
Dixmier-Douady classes for \emph{local data} of the gerbes. The canonical
$S^{1}$-bundle gerbe on the 3-sphere $S^{3}$ was first constructed
in \cite{Gawedzki1988} and later \cite{GawedzkiReis2002,Meinrenken2003}.
The canonical $U(1)$ - gerbe ${\cal {G}}$ on $S^{3}$ corresponds
to the 3-form $H=\frac{1}{12\pi}tr(g^{-1}dg)^{3}$. Other gerbes on
$S^{3}$, correspond to the curvatures $kH,\: k\in\mathbb{Z}$, and
are determined by the tensor powers ${\cal {G}}^{k}$. Given $kH,\, k\in\mathbb{Z}$
one has the unique gerbe ${\cal {G}}^{k}$ up to stable isomorphism,
since $H^{2}(SU(2),U(1))=\{1\}$.

Following the idea of Section \ref{sec:deformation-standard-R4},
we deform the standard $\mathbf{R}^{4}$ to get the small exotic $R^{4}$
together with the embedding $R^{4}\hookrightarrow\mathbf{R}^{4}$.
Then using Theorem \ref{thm:foliation-S3-from-Y} we obtain a codimension-one
foliation with non-zero Godbillon-Vey number on the 3-sphere in $\mathbf{R}^{4}$
represented by an element in $H^{3}(S^{3},\mathbb{R})$. Now we restrict
this value to the integers, i.e. we obtain an element in $H^{3}(S^{3},\mathbb{Z})$
corresponding to a foliation given by a flat $PSL(2,\mathbb{R})$
bundle (see the previous subsection). Every element in $H^{3}(S^{3},\mathbb{Z})$
determines an abelian gerbe with connection (and vice verse). Therefore
given a class in $H^{3}(S^{3},\mathbb{Z})$ we always determine a
corresponding gerbe with a connection and conversely, a gerbe representing
the class determines the foliation of $S^{3}\subset\mathbb{R}^{4}$
whose GV class agrees with those of the gerbe. Such foliation, however,
determines the (isotopy class of the embedding of) exotic $R^{4}$
in the 4-region bounded by $S^{3}$, as the member of the fixed radial
family. Thus, we have the result: \begin{theorem} \label{thm:exotic-R4-gerbes}
Every member $R_{t}^{4}$ of the fixed radial family embedded in $\mathbf{R}^{4}$
and represented by an integer Godbillon-Vey number, is generated from
an abelian gerbe with connection on $S^{3}$ in the standard $\mathbf{R}^{4}$.\par And
conversely: every member of the fixed radial family i.e. small exotic
$R^{4}$ together with an embedding $R^{4}\hookrightarrow\mathbf{R}^{4}$,
determines an abelian gerbe with a connection whose curvature class
is an integer Godbillon-Vey invariant. \end{theorem} Thus the effects
of the change of the smooth structure on $\mathbb{R}^{4}$ from the
standard one to any exotic structure (from the radial family) having
integer GV invariant, can be described by the corresponding twistings
of special structures on $S^{3}$, caused by the abelian gerbes on
$S^{3}$. In section \ref{sec:deformation-standard-R4} we studied
the effects on the geometry represented by the holonomy (as observables
of classical geometry) by deforming the standard $\mathbf{R}^{4}$
. We argued that the deformation changed the algebra of observables
from functions over holonomies to functions over knots/links. Using
our previous work \cite{AsselmeyerKrol2013}, this process is a deformation
quantization (also known as Turaev-Drinfeld quantization \cite{Turaev1991}).
Then a formal wave function is given by a section of a complex line
bundle over the space of (singular) knots/links. But this bundle is
a gerbe (see \cite{Brylinski1993} Theorem 6.3.1). Then the small
exotic $\mathbb{R}^{4}$ can be understood as a kind of quantized
geometry. This connection must be further studied in our forthcoming
work. Now we will concentrate on the case of non-integer Godbillon-Vey
numbers.

\section{Generalized (complex) structures of Hitchin \label{sub:Generalized-complex-str}}

The general non-integral case is naturally captured by the use of
generalized geometries of Hitchin. In this section we show the correspondence
between structures defined on $S^{3}$ whose variations reflect the
change of smoothness on $\mathbb{R}^{4}$. These structures are generalized
geometries and complex structures%
\footnote{Generalized complex structures require an even dimension of the manifold
hence should be considered on $S^{3}\times\mathbb{R}$. %
} introduced by Hitchin \cite{Hitchin1999} now called Hitchin structures.

Generalized structures are based on the substitution of the tangent
space $TM$ of a manifold $M$ by the sum $TM\oplus T^{\star}M$ of
the tangent and cotangent bundles such that the spin structure for
this generalized ,,tangent'' bundle becomes the bundle of all forms
$\wedge^{\bullet}M$ on $M$. Our interest in Hitchin structures was
arosed by the relation of deformations for this structure to the members
$R_{t}^{4}$ in the radial family of small exotic $R^{4}$s. We know
from Theorem \ref{thm:codim-1-foli-radial-fam} that the members $R_{t}^{4}$
are (at least) non-isotopic for different values $t\in[0,1]$. As
we will explain below, for these real values there exist corresponding
deformations of the Hitchin structure on $S^{3}$. Using the 2-framing
of a 3-manifold, we were able to motivate the appearance of Hitchin
structures above (see section \ref{sec:deformation-standard-R4}).
The theorem \ref{thm:correspondence-hitchin-exoticR4} will describe
the deformations of this Hitchin structure and their relation to small
exotic $\mathbb{R}^{4}$. But at first we have to introduce some notations.

The class of $H$- deformed Hitchin's structures on $S^{3}$ could
be as well referred to as \emph{$H$-twisted Courant brackets} on
$TS^{3}\oplus T^{\star}S^{3}$ and can be integrable with respect
to the brackets. In the following we will explain main points of this
correspondence which are mostly standard now. The advantage of using
generalized geometries is the apparent enhancing the diffeomorphisms
(of $M=S^{3}$) by the (non-trivial) $B$-field transformations. This
results in the $e^{B}$ extensions of the Courant bracket symmetries
corresponding to the change of the smoothness on $\mathbf{R}^{4}$
and is a way of passing from standard to nonstandard diffeomorphisms.
An excellent reference for generalized geometries and complex structures
is \cite{GualtieriPHD2005}. \begin{definition} Given a smooth manifold
$M$, the Courant bracket $[\,,\,]$ is defined on smooth sections
of $TM\oplus T^{\star}M$, by\par 
\begin{equation}
[X+\xi,Y+\eta]=[X,Y]+{\cal L}_{X}\eta-{\cal L}_{Y}\xi-\frac{1}{2}d(i_{X}\eta-i_{Y}\xi)\label{eq:CB}
\end{equation}
where $X+\xi$, $Y+\eta\in C^{\infty}(TM\oplus T^{\star}M)$, ${\cal L}_{X}$
is the Lie derivative in the direction of the field $X$, $i_{X}\eta$
is the inner product of a 1-form $\eta$ and a vector field $X$.
A Courant bracket on $M$ is also called a Hitchin structure or generalized
geometric (complex) structure \end{definition} On the RHS of (\ref{eq:CB})
$[\,,\,]$ is the Lie bracket on fields. This fact is not misleading
since the Courant bracket reduces to the Lie bracket for vector fields,
i.e. $\pi([X,Y])=[\pi(X),\pi(Y)]$ where $\pi:TM\oplus T^{\star}M\to TM$.
It follows that the bracket is skew symmetric and vanishes on 1-forms.
However, the Courant bracket is not a Lie bracket, since it does not
fulfill the Jacobi identity. The expression measuring the failure
of the identity is the so-called Jacobiator:

\begin{equation}
{\rm Jac}(X,Y,Z)=\left[[X,Y],Z\right]+\left[[Y,Z],X\right]+\left[[Z,X],Y\right]\label{eq:Jacob}
\end{equation}
The Jacobiator can be expressed as the derivative of a quantity which
is Nijenhuis operator, and it holds

\begin{equation}
{\rm Jac}(X,Y,Z)=d{\rm Nij}(X,Y,Z)\label{eq:Nij}
\end{equation}

\begin{equation}
{\rm Nij}(X,Y,Z)=\frac{1}{3}(\langle[X,Y],Z\rangle+\langle[Y,Z],X\rangle+\langle[Z,X],Y\rangle)\label{eq:Nij2}
\end{equation}
where $\langle\,,\,\rangle$ is the inner product on $TM\oplus T^{\star}M$.
This inner product is a naturally given by

\begin{equation}
\langle X+\xi,Y+\eta\rangle=\frac{1}{2}(\xi(Y)+\eta(X))\label{eq:prod<>}
\end{equation}
This product is symmetric and has the signature $(n,n)$, where $n={\rm dim}(M)$,
having the non-compact orthogonal group $O(TM\oplus T^{\star}M)=O(n,n)$
as symmetry of the product. \begin{definition} A subbundle $L<TM\oplus T^{\star}M$
is \emph{involutive} iff it is closed under the Courant bracket defined
on its smooth sections, and is \emph{isotropic }when $\langle X,Y\rangle=0$
for all $X,Y\in C^{\infty}(L)$ smooth sections of $L$. In case of
dim$(L)=n$ (maximality) we call the isotropic subbundle a \emph{maximal
isotropic subbundle. }\end{definition} The following property characterizes
these sub-bundles (see \cite{GualtieriPHD2005}, Proposition 3.27):

\emph{If $L$ is a maximal isotropic subbundle of $TM\oplus T^{\star}M$
then the following expressions are equivalent:} 
\begin{itemize}
\item \emph{$L$ is involutive} 
\item \emph{${\rm {Nij}_{L}=0}$} 
\item \emph{${\rm {Jac}_{L}}=0$.} 
\end{itemize}
\begin{definition} A Dirac structure\emph{ }on $TM\oplus T^{\star}M$
is a maximal isotropic and involutive subbundle\emph{ $L<TM\oplus T^{\star}M$.
}\end{definition} It follows from the properties above that the Dirac
structure is given by \emph{${\rm {Nij}_{L}=0}$}. The advantage of
using the Dirac structures is their generality - the structures used
in Poisson geometry, complex structures, foliated or symplectic geometries
are all special instances of Dirac structures. This Dirac structure
has therefore a great unifying power. The $H$- deformed Dirac structures
include also generalized complex structures which are well defined
on some manifolds without any complex or symplectic structures. Moreover,
this kind of geometry became extremely important in string theory
(flux compactification, mirror symmetry, branes in YM manifolds) and
related WZW models. As we will show now, these $H$- deformed Dirac
structures are also important for analyzing isotopy classes and of
exotic smoothness structures on $\mathbb{R}^{4}$. The main idea behind
this approach is the suitable modification of the Lie product of fields
on smooth manifolds.%
\footnote{Such a modification was suggested to one of the authors by Robert
Gompf some time ago.%
}

In differential geometry, the Lie bracket of smooth vector fields
on a smooth manifold $M$ is invariant under diffeomorphisms, and
there are no other symmetries of the tangent bundle preserving the
Lie bracket. More precisely, let $(f,F)$ be a pair of diffeomorphisms
$f:M\to M$ and $F:TM\to TM$ and $F$ is linear on each fiber, $\pi$
be the canonical projection $\pi:TM\to M$. Suppose that $F$ preserves
the Lie bracket $[\,,\,]$, i.e. $F([X,Y])=[F(X),F(Y)]$ for vector
fields $X$, $Y$ on $M$, and suppose the naturalness of $(f,F)$
i.e. $\pi\circ F=f\circ F$, then $F$ has to be equal to $df$.

In case of our extended ,,tangent space'' $TM\oplus T^{\star}M$,
the Courant bracket and the inner product are diffeomorphisms invariant.
However, there exists another symmetry extending the diffeomorphisms
which is the so-called $B$- field transformation. Let us see how
this work. Given a two-form $B$ on $M$ one can think of it as the
map $TM\to T^{\star}M$ by contracting $B$ with $X$, $X\to i_{X}B$.
Then the transformation of $TM\oplus T^{\star}M$ is given by $e^{B}:X+\xi\to X+\xi+i_{X}B$
with the properties (see \cite{GualtieriPHD2005}, Propositions 3.23,
3.24) 
\begin{itemize}
\item \emph{The map $e^{B}$ is an automorphism of the Courant bracket if
and only if $B$ is closed, i.e. $dB=0$,} 
\item \emph{The $e^{B}$extension of diffeomorphisms are the only allowed
symmetries of the Courant bracket.} 
\end{itemize}
It means, that for a pair $(f,F)$ consisting of a (orthogonal) automorphism
of $TM\oplus T^{\star}M$ and a transformation $F$ preserving the
Courant bracket $[\,,\,]$, i.e. $F([A,B])=[F(A),F(B)]$ for all sections
$A,B\in$$C^{\infty}(TM\to T^{\star}M)$. Therefore $F$ has to be
a composition of a diffeomorphism of $M$ and a $B$- field transform.
\emph{The group of orthogonal Courant automorphisms of $TM\oplus T^{\star}M$
is the semi-direct product of $Diff(M)$ and $\Omega_{closed}^{2}$.}

Given a Courant bracket on $TM\oplus T^{\star}M$ we are able to define
various involutive structures by using this bracket. The most important
possibility is the deformation of the Courant bracket on $TM\oplus T^{\star}M$
by a real closed 3-form $H$ on $M$. For any real 3-form $H$ one
has the twisted Courant bracket on $TM\oplus T^{\star}M$ defined
as

\begin{equation}
[X+\xi,Y+\eta]_{H}=[X+\xi,Y+\eta]+i_{Y}i_{X}H\label{eq:HtwistedCourantBracket}
\end{equation}
where {[}$\,,\,${]} on the RHS is the non-twisted Courant bracket.
This can be also restated as the splitting condition in the non-trivial
twisted Courant algebroid defined later.

This deformed bracket allows to define various involutive and (maximal)
isotropic structures with respect to $[\,,\,]_{H}$, which is again
an analog for the integrability of distributions on manifolds. These
structures correspond to new $H$-twisted geometries which are different
for the previously considered Dirac structures in case of the untwisted
Courant bracket. In particular, the $B$- field transform of $[\,,\,]_{H}$
is the symmetry of the bracket if and only if $dB=0$, since it yields

\begin{equation}
\left[e^{B}(C),e^{B}(D)\right]_{H}=e^{B}\left[C,D\right]_{H+dB},\label{eq:Btrans.HdeformCB}
\end{equation}
for all $C,D\in C^{\infty}(TM\oplus T^{\star}M)$. Then the tangent
bundle $TM$ is not involutive with respect to $[\,,\,]_{H}$ for
non-zero $H.$ In general a subbundle $L$ is involutive with respect
to $[\,,\,]_{H}$ if and only if $e^{-B}L$ is involutive for $[\,,\,]_{H+dB}$.

In the following we will comment on the relation between Hitchin structures
and gerbes used in the next theorem. Furthermore we are able to understand
a way how $TM\oplus T^{\star}M$ appears from the broader perspective
of the extensions of bundles. For that purpose one defines the \emph{Courant
algebroid $E$} as an extension of real vector bundles given by the
sequence

\begin{equation}
0\to T^{\star}M\to_{\pi^{\star}}E\to_{\pi}TM\to0\label{eq:Eextension}
\end{equation}
On $E$ a non-degenerate symmetric bilinear form $\langle\,,\,\rangle$
is given, such that $\langle\pi^{\star}\xi,a\rangle=\xi(\pi(a))$
where $\xi$ is smooth covector field on $M$ and $a\in C^{\infty}(E)$
. A bilinear Courant bracket $[,]$ on $C^{\infty}(E)$ can be defined
such that 
\begin{itemize}
\item $[a,[b,c]]-[[a,b],c]+[b,[a,c]]=0$ (Jacobi identity) 
\item $[a,fb]=f[a,b]+(\pi(a)f)b$ (Leibniz rule) 
\item $\pi(a)\langle b,c\rangle=\langle[a,b],c\rangle+\langle b,[a,c]\rangle$
(Invariance of bilinear form) 
\item $[a,a]=\pi^{\star}d\langle a,a\rangle$ 
\end{itemize}
The sequence (\ref{eq:Eextension}) defines a splitting of $E$. Each
splitting determines a closed 3-form $H\in\Omega^{3}(M)$, given by

\begin{equation}
(i_{X}i_{Y}H)(Z)=\langle[s(X),s(Y)],s(Z)\rangle\label{eq:Hinsplitting}
\end{equation}
where $s:TM\to E$ is the splitting derived from the sequence. The
cohomology class $[H]/2\pi\in H^{3}(M,\mathbb{R})$ is independent
of the choice of splitting, and coincides with the image of the Dixmier-Douady
class of the gerbe in real cohomology \cite{Hitchin2005}. The Dixmier-Douady
class classifies the bundle gerbes as done in Section \ref{sub:U(1)--gerbes-on}.

\begin{theorem} \emph{\label{thm:correspondence-hitchin-exoticR4}}
Given a radial family of small exotic $\mathbb{R}^{4}$ and radius
$r$, the Godbillon-Vey invariant is an element of $H^{3}(S^{3},\mathbb{R})\thickapprox\mathbb{R}$
and $GV=r^{2}$. Every small exotic $\mathbb{R}^{4}$, called $R^{4}$,
of the fixed radial family admits an embedding $R^{4}\hookrightarrow\mathbf{R}^{4}$.
There is a foliated cobordism between the 3-sphere $S^{3}\subset\mathbf{R}^{4}$
and the 3-manifold $\Sigma=Y_{n}\subset R^{4}$ separating a compact
subset of the exotic $\mathbb{R}^{4}$ from infinity. Then, there
are the following correspondences: \begin{itemize} 

\item A. i. The continuous family of distinct small exotic smooth
structures on $\mathbb{R}^{4}$ corresponds to the $H$-deformed classes
of generalized Hitchin's geometries on $S^{3}$, where $[H]\in H^{3}(S^{3},\mathbb{R})$.\par ii.
Further they correspond $1\div1$ to the family of deformations of
the Hitchin geometry on the 3-manifold $\Sigma=Y_{n}$ by $[H_{\Sigma}]\in H^{3}(\Sigma,\mathbb{R})$.\par iii.
Moreover, all deformations $[H]\in H^{3}(S^{3},\mathbb{R})$ can be
realized by a noncobordant codimension-one foliations of the 3-sphere
$S^{3}$. These foliations are determined by the small exotic $\mathbb{R}^{4}$s
from the radial family. Each such foliation defines a Dirac structure. 

\item B. For integral classes $[H]\in H^{3}(S^{3},\mathbb{Z})$,
the deformations are geometrically described by $S^{1}$ gerbes. (see
also theorem \ref{thm:exotic-R4-gerbes}) \end{itemize} \end{theorem}\emph{
Proof:} We start to show the correspondence A. The class $[H]\in H^{3}(S^{3},\mathbb{R})$
deforming the Courant bracket on $TS^{3}\oplus T^{\star}S^{3}$ will
be used to relate it with the small exotic $\mathbb{R}^{4}$ by theorem
\ref{thm:codim-1-foli-radial-fam}. Because of the isomorphism $S^{3}=SU(2),$
the tangent bundle $TS^{3}$ is trivial, i.e. $TS^{3}=S^{3}\times\mathbb{R}^{3}$
and there are three global vector fields (as sections) generating
(locally) every other vector field by linear combination. The triviality
of $H^{2}(S^{3},\mathbb{R})=0$ implies that $\Omega_{closed}^{2}$
are given by all exact 2-forms as image of all 1-forms $\Omega_{closed}^{2}=d\Omega^{1}$.
Therefore the automorphisms are mainly generated by the diffeomorphism
$Diff(S^{3})$. Now we consider the 1-form $\omega$ defining the
codimension-one foliation on $S^{3}$ with the dual $X=\omega^{*}$
, a vector field. This vector field generates a 1-dimensional subbundle
$\triangle\subset TS^{3}$ so that $\triangle\oplus Ann(\triangle)\subset TS^{3}\oplus T^{*}S^{3}$
($Ann()$ is the annihilator) is a maximal isotropic subbundle, hence
defines a Dirac structure proving last part of iii. (see \cite{GualtieriPHD2005}
Example 3.32). Now we choose the closed 3-form $H=\Gamma_{\mathcal{F}}$
with the Godbillon-Vey class $\Gamma_{\mathcal{F}}$ as deformation
of the Hitchin structure. We know from theorem \ref{thm:codim-1-foli-radial-fam}
that the radial family of small exotic $\mathbb{R}^{4}$'s is determined
by the noncobordant codimension-one foliations of $\Sigma$ and by
the theorems \ref{thm:foliation-3MF} and \ref{thm:foliation-S3-from-Y}
it is also determined by noncobordant codimension-one foliations of
$S^{3}$ as well. The discussion can also extended to the 3-manifold
$\Sigma$ showing the full correspondence A. Let us mention the possibility
to take generalized complex structures on $S^{3}\times\mathbb{R}$
by using the isomorphism $H^{3}(S^{3}\times\mathbb{R},\mathbb{R})\simeq H^{3}(S^{3},\mathbb{R})$
by considering the end of $\mathbf{R}^{4}$.

To show the correspondence B, we will use the comments about the Courant
algebroid above. The condition (\ref{eq:gerbeConnection1}) in Sec.
\ref{sub:U(1)--gerbes-on} for the $S^{1}$-gerbe with connection
has the interpretation that $dA_{\alpha\beta}$ is a cocycle. Using
this interpretation, one can locally paste together $(TM\oplus T^{\star}M)_{\alpha}$
and $(TM\oplus T^{\star}M)_{\beta}$ by the automorphism $\left(\begin{array}{cc}
1 & 0\\
dA_{\alpha\beta} & 1
\end{array}\right)$. The action of $dA_{\alpha\beta}$ on $TM$ is defined by $X\to i_{X}dA_{\alpha\beta}$.
The second condition for the connection data of a gerbe (see (\ref{eq:gerbeConnection2})
in Sec. \ref{sub:U(1)--gerbes-on}) defines a splitting of the Courant
algebroid like in (\ref{eq:Hinsplitting}). Then we can make use of
the proposition 3.47 in \cite{GualtieriPHD2005} ( \emph{If $[H/2\pi]\in H^{3}(M,\mathbb{Z})$
then the twisted Courant bracket $[,]_{H}$ on $TM\oplus T^{\star}M$
can be obtained from a $S^{1}$ gerbe with connection.}) If $[H/2\pi]$
is integral, then because of $dB=0$ and (\ref{eq:Btrans.HdeformCB})
the trivializations $f_{\alpha\beta}$ of a \emph{flat gerbe with
connection} are symmetries of $[,]_{H}$ (B-field transforms). The
difference of two possible trivializations is a line bundle with connection,
see Sec. \ref{sub:U(1)--gerbes-on}. These line bundles have the role
of \emph{gauge transformations} (integral B-fields) \cite{GualtieriPHD2005}
establishing the correspondence B. $\square$

A general result for the family of small smooth $\mathbb{R}^{4}$'s
can be restated as\\
 \emph{The change of exotic smooth structure on $\mathbb{R}^{4}$
results in the change of a generalized Dirac structures on $S^{3}$
lying in the standard $\mathbf{R}^{4}$}.

\section{Charge quantization without magnetic monopoles}

The connection between the embedding classes of some small exotic
$\mathbb{R}^{4}$ in $\mathbf{R}^{4}$ constructed from the radial
family and the codimension-one foliations of $S^{3}$ with Godbillon-Vey
classes represented by 3-rd cohomologies $H^{3}(S^{3},\mathbb{R})$,
enables one to realize geometrical effects of these exotica in $\mathbf{R}^{4}$
via abelian gerbes on $S^{3}$ and twisted Courant brackets. The direct
relation between the integral classes in $H^{3}(S^{3},\mathbb{Z})$
and exotic $\mathbb{R}^{4}$ was discussed in subsection \ref{sub:Integer-Godbillon-Vey-invariants}.
In this section we will discuss this relation as a kind of localization
principle causing strong physical consequences. Namely, the condition
for magnetic monopoles in spacetime is expressed in terms of abelian
gerbes which gives non-vanishing third integral cohomologies $H^{3}(S^{3},\mathbb{Z})$
(\cite{Brylinski1993}, Chapter 7). As we saw in Sec. \ref{sec:Exotic-R4-codim-1-foliation}
the change of smoothness on $\mathbb{R}^{4}$ between the members
of the radial family for integer radii, corresponds to the change
between the corresponding integral classes $H^{3}(S^{3},\mathbb{Z})$.
Here we will state the stronger \emph{localization principle}:

\emph{Effects of exotic smooth $\mathbb{R}^{4}$ in a region of the
4-spacetime $M^{4}$ can be localized in $S^{3}$ embedded in $M^{4}$,
i.e. the appearance of this exotic smoothness is equivalent to the
appearance of the nontrivial, geometrically realized class in $H^{3}(S^{3},\mathbb{Z})$.
This class is stable with respect to continuous deformations.}

Here we will discuss the physical consequence of the exoticness of
an open 4-region in 4-spacetime:

\emph{The quantization condition for electric charge in spacetime
can be seen as a consequence of certain non-standard 4-smoothness
appearing in spacetime.} \\
 In the following we will present the details to understand this consequence.

\subsection{Dirac's magnetic monopoles and $S^{1}$ - gerbes }

When the magnetic field $B$ is defined over the whole euclidean space
$\mathbb{R}^{3}$, there exists a globally defined vector potential,
and any two vector potentials differ by the gradient of some function.
In terms of the connection on the line bundle $L$, one can trivialize
the bundle, and the connection $\nabla$ on $L$ is given by

\begin{equation}
\nabla=d+i\frac{e}{\hbar c}A\label{eq:connection1}
\end{equation}
Dirac considered a magnetic field $B$ defined on $\mathbb{R}^{3}\setminus\left\{ 0\right\} $
which has a singularity at the origin. This singularity corresponds
to the existence of a magnetic monopole localized at the origin. The
magnetic monopole has the strength $\mu$

\begin{equation}
\mu=\frac{1}{4\pi}\int\intop_{S^{2}}\overrightarrow{B}\times d\sigma\label{eq:strength}
\end{equation}
which is the flux of $\overrightarrow{B}$ through the 2-sphere $S^{2}$
up to a constant. Because of $div(\overrightarrow{B})=0$, the integral
does not depend on the choice of the $\Sigma$ centered at the origin.

Equivalently $\mu$ can be expressed in terms of the curvature 2-form
$R$, of a connection on $L$, as

\begin{equation}
\mu=-i\frac{c\hbar}{4\pi e}\intop_{S^{2}}R\label{eq:strength-curvature}
\end{equation}
Now, the integral $\intop_{S^{2}}R$ has values which are integer
multiplicities of $2\pi i$ on a complex line bundle $L$. Thus, the
following quantization condition for the strength of a magnetic monopole,
follows

\begin{equation}
\mu=\frac{c\hbar}{2e}\cdot n,\: n\in\mathbb{Z}\label{eq:quantized-strength}
\end{equation}

The whole discussion based on the fact that the cohomology class of
$\frac{R}{2\pi i}$ is integral and the magnetic field $\overrightarrow{B}$
is proportional to the curvature of some line bundle with connection
on $\mathbb{R}^{3}\setminus\left\{ 0\right\} $. We see, that (\ref{eq:quantized-strength})
is the same as $\frac{2}{c\hbar}\mu\cdot e=n$ and this means that
the electrical charge has to be quantized.

The cohomologies involved here are $H^{2}(\mathbb{R}^{3}\setminus\left\{ 0\right\} )$.
We extend, following \cite{Brylinski1993}, Chap. 7, the forms and
cohomologies over the whole 3-space, including the origin. To this
end let us consider generalized 3-forms on $\mathbb{R}^{3}$ which
are supported at the origin, i.e. the relative cohomology group $H^{3}(\mathbb{R}^{3},\mathbb{R}^{3}\setminus\left\{ 0\right\} )=:H_{0}^{3}(\mathbb{R}^{3})$.
Given the exact sequence

\begin{equation}
H^{2}(\mathbb{R}^{3})=0\to H^{2}(\mathbb{R}^{3}\setminus\left\{ 0\right\} )\to H_{0}^{3}(\mathbb{R}^{3})\to H^{3}(\mathbb{R}^{3})=0\label{eq:exact-seq}
\end{equation}
we have the isomorphism

\begin{equation}
H^{2}(\mathbb{R}^{3}\setminus\left\{ 0\right\} )=H_{0}^{3}(\mathbb{R}^{3})\label{eq:homology3/2}
\end{equation}

We saw in (\ref{eq:quantized-strength}) that a topological analog
of the monopole is the element of the 2-nd cohomology $H^{2}(\mathbb{R}^{3}\setminus\left\{ 0\right\} ,\mathbb{Z})$.
Thus, the extension of the description of a monopole, located at the
origin, to the entire $\mathbb{R}^{3}$, gives the topological analog
of the monopole as an element of $H_{0}^{3}(\mathbb{R}^{3})$. A monopole
is moving now inside the 3-space, and from the canonical isomorphisms
(see \cite{Brylinski1993} formulas (7-23) and (7-24) on page 264)

\begin{equation}
H_{0}^{3}(\mathbb{R}^{3})\simeq H_{0}^{3}(\mathbb{R}^{3}\cup\left\{ \infty\right\} )\simeq H^{3}(S^{3})\label{eq:isomorphisms}
\end{equation}
the topological counterpart of a monopole is the element of $H^{3}(S^{3},\mathbb{Z})$.
This isomorphism can be also written in the form $H^{2}(S^{2})\simeq H^{3}(S^{3})$
(using (\ref{eq:homology3/2}) and $H^{2}(\mathbb{R}^{3}\setminus\left\{ 0\right\} )=H^{2}(S^{2})$)
and we will show it by another method. The cohomology of a manifold
$M$ is also defined by $H^{n}(M,G)=[M,K(n,G)]$ with the Eilenberg-MacLane
space $K(n,G)$ (for arbitrary coefficients $G$). Now there is the
suspension functor $\Sigma_{*}$ so that $\Sigma_{*}S^{2}$ is homotopic
to $S^{3}$. Then one obtains 
\[
H^{3}(S^{3},G)=[S^{3},K(3,G)]\simeq[\Sigma_{*}S^{2},K(3,G)]\simeq[S^{2},\Omega K(3,G)]=H^{2}(S^{2},G)
\]
with the loop functor $\Omega$ obeying the property $[\Sigma_{*},\,]\simeq[\,,\Omega]$
and using the homotopy equivalence between $\Omega K(n,G)$ and $K(n-1,G)$.
This isomorphism is canonical by construction (see also \cite{Spa:66}).
Conversely, the topological realization of some element in $H^{3}(S^{3},\mathbb{Z})$
is given by some line bundle with connection on $\mathbb{R}^{3}\setminus\left\{ 0\right\} $,
which is equivalent to the existence of a Dirac monopole in spacetime
and consequently the electrical charge is quantized.

The whole discussion can be extended to the relativistic theory in
$\mathbb{R}^{4}$ as well (see the introduction of \cite{BoTu:82}).
Then we consider the Coulomb potential of a point particle of charge
$q$ in $0\in\mathbb{R}^{4}$ as the connection one-form of a line
bundle 
\[
A=-q\cdot\frac{1}{r}\cdot dt
\]
over $\mathbb{R}^{4}\setminus\mathbb{R}_{t}$ with $r^{2}=x^{2}+y^{2}+z^{2}\not=0$.
The curvature is given by the exact 2-form 
\[
F=dA
\]
fulfilling the first Maxwell equation $dF=0$. Now we consider the
2-form $*F$ 
\[
*F=\frac{q}{4\pi}\cdot\frac{x\, dy\wedge dz-y\, dx\wedge dz+z\, dx\wedge dy}{r^{3}}
\]
as element of $H^{2}(\mathbb{R}^{4}\setminus\mathbb{R}_{t})$ fulfilling
the second Maxwell equation $d*F=0$. A simple argument showed the
isomorphism 
\[
H^{*}(\mathbb{R}^{4}\setminus\mathbb{R}_{t})\cong H^{*}(\mathbb{R}^{3}\setminus\left\{ 0\right\} )
\]
and the integral along an embedded $S^{2}$ surrounding the point
$0$ gives 
\[
\intop_{S^{2}}*F=q
\]
the charge $q$ of the particle. Together with the isomorphism (\ref{eq:homology3/2})
we obtain the relation between a relativistic particle of charge $q$
in $\mathbb{R}^{4}$ and elements of $H^{3}(S^{3})$.

The elements of $H^{3}(S^{3},\mathbb{Z})$ have well-defined topological
realizations like the elements of $H^{2}(\mathbb{R}^{3}\setminus\left\{ 0\right\} ,\mathbb{Z})$
corresponding to line bundles with connection. Namely, $h\in H^{3}(S^{3},\mathbb{Z})$
corresponds to the $S^{1}$- gerbe ${\cal G}_{h}$ on $S^{3}$. However,
the elements of $H^{3}(S^{3},\mathbb{Z})$ have yet another realizations
in spacetime.

\subsection{$S^{1}$-gerbes on $S^{3}$ and exotic smooth $R^{4}$'s}

In Sec. \ref{sec:Exotic-R4-codim-1-foliation} it was shown that third
real deRham cohomology classes of $S^{3}$ correspond to the isomorphism
classes of codimension-one foliations of $S^{3}$ (see Theorem \ref{thm:codim-1-foli-radial-fam}).
For integral 3-rd cohomologies of $S^{3}$ we have the correspondence
between $S^{1}$- gerbes on $S^{3}$ and the isotopy classes of embeddings
of some exotic smooth $R^{4}$s in $\mathbf{R}^{4}$. The correspondence
means, in particular, that some small exotic smooth $R^{4}$ embedded
in some open region in $\mathbf{R}^{4}$ serves as the spacetime realization
of the integral 3-rd cohomology class of $S^{3}$. In other words,
a non-trivial 3-rd cohomology class of $S^{3}$ is realized in spacetime
by exotic smooth 4-structure in some region. On the other hand, a
class in $H^{3}(S^{3},\mathbb{Z})$ (canonically isomorphic to $H^{2}(S^{2},\mathbb{Z})\simeq H^{2}(\mathbb{R}^{3}\setminus\left\{ 0\right\} ,\mathbb{Z})$)
is realized geometrically as a line bundle with connection on $\mathbb{R}^{3}\setminus\left\{ 0\right\} $.
From the point of physics in 4-spacetime, the existence of this line
bundle could be equivalent to the action of a monopole existing somewhere
in spacetime. However, if a monopole exists, electric charge is quantized.
We have to make some additional assumptions to be sure that exoticness
of a region of spacetime can give the same effect as the existence
of a monopole. Namely, we suppose that the magnetic field propagates
over a region with an exotic smoothness structure in 4-spacetime with
Minkowski metric such that 
\begin{itemize}
\item the exotic smooth $R_{h}^{4}$ corresponds to the class $[h]\in H^{3}(S^{3},\mathbb{Z})$
and 
\item the strength of the magnetic field is proportional to the curvature
of the line bundle on $\mathbb{R}^{3}\setminus\left\{ 0\right\} $
which corresponds to $[h]\in H^{2}(\mathbb{R}^{3}\setminus\left\{ 0\right\} )$. 
\end{itemize}
Then this exotic $R_{h}^{4}$ acts as a source for the magnetic field
in $\mathbb{R}^{4}$, i.e. the magnetic charge of the monopole and
the electric charge are quantized, see (\ref{eq:quantized-strength}).
Given a class $[h]\in H^{2}(\mathbb{R}^{3}\setminus\left\{ 0\right\} )$
we always find a monopole solution fulfilling these requirements.
By (\ref{eq:quantized-strength}), a monopole gives the quantization
of the electric charge and is the source for the magnetic field. As
remarked in subsection \ref{sub:Integer-Godbillon-Vey-invariants},
the codimension-one foliation leading to integer Godbillon-Vey numbers
is stable with respect to deformations. Therefore, the quantization
is stable with respect to continuous deformations as well.

The \emph{Brans conjecture} \cite{Bra:94a} states that exotic smooth
structures on 4-manifolds (compact and non-compact) can serve as external
sources for the gravitational field \cite{Bra:94b}. This conjecture
was proven by Asselmeyer \cite{Ass:96} in the compact case, and by
S{\l{}}adkowski \cite{Sladkowski2001} in the non-compact case and
for large exotic $R^{4}$. Here we yield an essential extension of
the Brans conjecture for magnetic fields:\\
 \emph{Some small, exotic smooth structures on $\mathbb{R}^{4}$ embedded
in the standard $\mathbf{R}^{4}$ can act as sources of magnetic field,
i.e. monopoles, in spacetime. The electrical charge in spacetime has
to be quantized, provided some region has this small exotic smoothness.}\\
 The connection of small exotic $\mathbb{R}^{4}$ with magnetism and
spins of particles is especially important and opens a wide range
of further physical applications (see e.g. \cite{AsselmeyerKrolProc2014}).

\section*{Acknowledgment}

We thank L. Taylor for the discussion of an error in our argumentation.
Special thanks to R. Gompf for a long email discussion where many
points were clarified. T.A. wants to thank C.H. Brans and H. Ros{é}
for numerous discussions over the years about the relation of exotic
smoothness to physics. J.K. benefited much from the explanations given
to him by R. Gompf regarding 4-smoothness several years ago, and discussions
with J. S{\l{}}adkowski.

Last but not least we acknowledged all critical remarks of the referees
increasing strongly the readability of the paper.

\section*{Appendix A Casson handles and labeled trees}

Given a labeled based tree $Q$, let us describe a subset $U_{Q}$
of $D^{2}\times D^{2}$. Now we will construct a $(U_{Q},\partial D^{2}\times D^{2})$
which is diffeomorphic to the Casson handle associated to $Q$. In
$D^{2}\times D^{2}$ embed a ramified Whitehead link with one Whitehead
link component for every edge labeled by $+$ leaving {*} and one
mirror image Whitehead link component for every edge labeled by $-$(minus)
leaving {*}. Corresponding to each first level node of $Q$ we have
already found a (normally framed) solid torus embedded in $D^{2}\times\partial D^{2}$.
In each of these solid tori embed a ramified Whitehead link, ramified
according to the number of $+$ and $-$ labeled branches leaving
that node. We can do that process for every level of $Q$. Let the
disjoint union of the (closed) solid tori in the $n$th family (one
solid torus for each branch at level $n$ in $Q$) be denoted by $X_{n}$.
$Q$ tells us how to construct an infinite chain of inclusions: 
\[
\ldots\subset X_{n+1}\subset X_{n}\subset X_{n-1}\subset\ldots\subset X_{1}\subset D^{2}\times\partial D^{2}
\]
and we define the Whitehead decomposition $WhC_{Q}=\bigcap_{n=1}^{\infty}X_{n}$
of $Q$. $WhC_{Q}$ is the Whitehead continuum \cite{Whitehead35}
for the simplest unbranched tree. We define $U_{Q}$ to be 
\[
U_{Q}=D^{2}\times D^{2}\setminus(D^{2}\times\partial D^{2}\cup\mbox{closure}(WhC_{Q}))
\]
alternatively one can also write 
\begin{equation}
U_{Q}=D^{2}\times D^{2}\setminus\mbox{cone}(WhC_{Q})\label{eq:UQ-diffeo-CH}
\end{equation}
where $\mbox{cone}()$ is the cone of a space 
\[
cone(A)=A\times[0,1]/(x,0)\sim(x',0)\qquad\forall x,x'\in A
\]
over the point $(0,0)\in D^{2}\times D^{2}$. As Freedman (see \cite{Fre:82}
Theorem 2.2) showed $U_{Q}$ is diffeomorphic to the Casson handle
$CH_{Q}$ given by the tree $Q$.

\section*{Appendix B 3-manifolds and geometric structures}

\begin{figure}
\centerline{\psfig{file=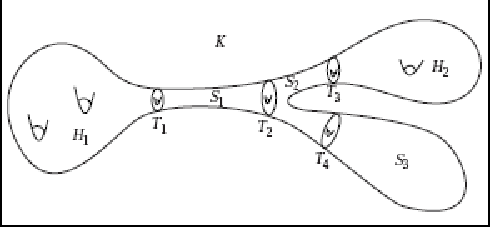,width=8cm}}
\caption{Torus (JSJ-) decomposition, $H_{i}$ hyperbolic manifold, $S_{i}$
Graph-manifold, $T_{i}$ Tori \label{fig:Torus-decomposition}}
\end{figure}

A connected 3-manifold $N$ is prime if it cannot be obtained as a
connected sum of two manifolds $N_{1}\#N_{2}$ neither of which is
the 3-sphere $S^{3}$ (or, equivalently, neither of which is the homeomorphic
to $N$). Examples are the 3-torus $T^{3}$ and $S^{1}\times S^{2}$
but also the Poincare sphere. According to \cite{Mil:62}, any compact,
oriented 3-manifold is the connected sum of an unique (up to homeomorphism)
collection of prime 3-manifolds (prime decomposition). A subset of
prime manifolds are the irreducible 3-manifolds. A connected 3-manifold
is irreducible if every differentiable submanifold $S$ homeomorphic
to a sphere $S^{2}$ bounds a subset $D$ (i.e. $\partial D=S$) which
is homeomorphic to the closed ball $D^{3}$. The only prime but reducible
3-manifold is $S^{1}\times S^{2}$. For the geometric properties (to
meet Thurstons geometrization theorem) we need a finer decomposition
induced by incompressible tori. A properly embedded connected surface
$S\subset N$ is called 2-sided%
\footnote{The `sides' of $S$ then correspond to the components of the complement
of $S$ in a tubular neighborhood $S\times[0,1]\subset N$.%
} if its normal bundle is trivial, and 1-sided if its normal bundle
is nontrivial. A 2-sided connected surface $S$ other than $S^{2}$
or $D^{2}$ is called incompressible if for each disk $D\subset N$
with $D\cap S=\partial D$ there is a disk $D'\subset S$ with $\partial D'=\partial D$.
The boundary of a 3-manifold is an incompressible surface. Most importantly,
the 3-sphere $S^{3}$, $S^{2}\times S^{1}$ and the 3-manifolds $S^{3}/\Gamma$
with $\Gamma\subset SO(4)$ a finite subgroup do not contain incompressible
surfaces. The class of 3-manifolds $S^{3}/\Gamma$ (the spherical
3-manifolds) include cases like the Poincare sphere ($\Gamma=I^{*}$
the binary icosaeder group) or lens spaces ($\Gamma=\mathbb{Z}_{p}$
the cyclic group). Let $K_{i}$ be irreducible 3-manifolds containing
incompressible surfaces then we can $N$ split into pieces (along
embedded $S^{2}$) 
\begin{equation}
N=K_{1}\#\cdots\#K_{n_{1}}\#_{n_{2}}S^{1}\times S^{2}\#_{n_{3}}S^{3}/\Gamma\,,\label{eq:prime-decomposition}
\end{equation}
where $\#_{n}$ denotes the $n$-fold connected sum and $\Gamma\subset SO(4)$
is a finite subgroup. The decomposition of $N$ is unique up to the
order of the factors. The irreducible 3-manifolds $K_{1},\ldots,\, K_{n_{1}}$
are able to contain incompressible tori and one can split $K_{i}$
along the tori into simpler pieces $K=H\cup_{T^{2}}G$ \cite{JacSha:79}
(called the JSJ decomposition). The two classes $G$ and $H$ are
the graph manifold $G$ and the hyperbolic 3-manifold $H$ (see Fig.
\ref{fig:Torus-decomposition}). 

The hyperbolic 3-manifold $H$ has a torus boundary $T^{2}=\partial H$,
i.e. $H$ admits a hyperbolic structure in the interior only. In this
paper we need the splitting of the link/knot complement. As shown
in \cite{Budney2006}, the Whitehead double of a knot leads to JSJ
decomposition of the complement into the knot complement and the complement
of the Whitehead link (along one torus boundary of the Whitehead link
complement).

One property of hyperbolic 3-manifolds is central: Mostow rigidity.
As shown by Mostow \cite{Mos:68}, every hyperbolic $n-$manifold
$n>2$ with finite volume has this property: \emph{Every diffeomorphism
(especially every conformal transformation) of a hyperbolic $n-$manifold
with finite volume is induced by an isometry.} Therefore one cannot
scale a hyperbolic 3-manifold and the volume is a topological invariant.
Together with the prime and JSJ decomposition 
\[
N=\left(H_{1}\cup_{T^{2}}G_{1}\right)\#\cdots\#\left(H_{n_{1}}\cup_{T^{2}}G_{n_{1}}\right)\#_{n_{2}}S^{1}\times S^{2}\#_{n_{3}}S^{3}/\Gamma\,,
\]
we can discuss the geometric properties central to Thurstons geometrization
theorem: \emph{Every oriented closed prime 3-manifold can be cut along
tori (JSJ decomposition), so that the interior of each of the resulting
manifolds has a geometric structure with finite volume.} Now, we have
to clarify the term geometric structure's. A model geometry is a
simply connected smooth manifold $X$ together with a transitive action
of a Lie group $G$ on $X$ with compact stabilizers. A geometric
structure on a manifold $N$ is a diffeomorphism from $N$ to $X/{\Gamma}$
for some model geometry $X$, where $\Gamma$
is a discrete subgroup of $G$ acting freely on $X$. t is a surprising
fact that there are also a finite number of three-dimensional model
geometries, i.e. 8 geometries with the following models: spherical
$(S^{3},O_{4}(\mathbb{R}))$, Euclidean $(\mathbb{E}^{3},O_{3}(\mathbb{R})\ltimes\mathbb{R}^{3})$,
hyperbolic $(\mathbb{H}^{3},O_{1,3}(\mathbb{R})^{+})$, mixed spherical-Euclidean
$(S^{2}\times\mathbb{R},O_{3}(\mathbb{R})\times\mathbb{R}\times\mathbb{Z}_{2})$,
mixed hyperbolic-Euclidean $(\mathbb{H}^{2}\times\mathbb{R},O_{1,3}(\mathbb{R})^{+}\times\mathbb{R}\times\mathbb{Z}_{2})$
and 3 exceptional cases called $\tilde{SL}_{2}$ (twisted version
of $\mathbb{H}^{2}\times\mathbb{R}$), NIL (geometry of the Heisenberg
group as twisted version of $\mathbb{E}^{3}$), SOL (split extension
of $\mathbb{R}^{2}$ by $\mathbb{R}$, i.e. the Lie algebra of the
group of isometries of 2-dimensional Minkowski space). We refer to
\cite{Scott1983} for the details.



\end{document}